\documentclass[fleqn,usenatbib]{mnras}

\usepackage{newtxtext,newtxmath}
\usepackage{threeparttable}
\usepackage[T1]{fontenc}
\usepackage{ae,aecompl}

\usepackage{graphicx}	% Including figure files
\usepackage{amsmath}	% Advanced maths commands
\usepackage{amssymb}	% Extra maths symbols
\usepackage{adjustbox}
\usepackage{rotating}
\usepackage{soul}

\title{Complex organic molecules in the Galactic Centre: the N-bearing family}

\author[S.Zeng et al.]{
S. Zeng,$^{1}$\thanks{E-mail: z.shaoshan@qmul.ac.uk}
I. Jim\'enez-Serra,$^{1}$
V. M. Rivilla,$^{2}$
S. Mart\'in,$^{3,4}$
J. Mart\'in-Pintado,$^{5}$
\newauthor{
M. A. Requena-Torres,$^{6}$
J. Armijos-Abenda\~no,$^{7}$
D. Riquelme,$^{8}$
and R. Aladro,$^{8}$}
\\
$^{1}$School of Physics and Astronomy, Queen Mary University of London, Mile End Road, E1 4NS London, United Kingdom\\
$^{2}$INAF-Osservatorio Astrofisico di Arcetri, Largo Enrico Fermi 5, I-50125, Florence, Italy\\
$^{3}$ European Southern Observatory (ESO), Alonso de C\'ordova 3107, Vitacura, Santiago, Chile\\
$^{4}$Joint ALMA Observatory, Alonso de C\'ordova 3107, Vitacura, Santiago, Chile\\
$^{5}$Centro de Astrobiolog\'ia (CSIC,INTA). Ctra de Ajalvir, km. 4, Torrej\'on de Ardoz, 28850 Madrid, Spain \\
$^{6}$University of Maryland, College Park, ND 20742-2421 \\
$^{7}$Observatorio Astron\'omico de Quito, Escuela Polit\'ecnica Nacional, Av. Gran Colombia S/N, Interior del Parque La Alameda,\\
       170136, Quito, Ecuador \\
$^{8}$Max-Planck-Institut f\"ur Radioastronomie, Auf dem H\"ugel 69, 53121 Bonn, Germany \\
}

\date{Accepted XXX. Received YYY; in original form ZZZ}

\pubyear{2018}

\begin{document}
\label{firstpage}
\pagerange{\pageref{firstpage}--\pageref{lastpage}}
\maketitle
\raggedbottom

\begin{abstract}
We present an unbiased spectral line survey toward the Galactic Centre (GC) quiescent giant molecular cloud (QGMC), G+0.693 using the GBT and IRAM 30$\,$ telescopes. Our study highlights an extremely rich organic inventory of abundant amounts of nitrogen (N)-bearing species in a source without signatures of star formation. We report the detection of 17 N-bearing species in this source, of which 8 are complex organic molecules (COMs). A comparison of the derived abundances relative to H$_2$ is made across various galactic and extragalactic environments. We conclude that the unique chemistry in this source is likely to be dominated by low-velocity shocks with X-rays/cosmic rays also playing an important role in the chemistry. Like previous findings obtained for O-bearing molecules, our results for N-bearing species suggest a more efficient hydrogenation of these species on dust grains in G+0.693 than in hot cores in the Galactic disk, as a consequence of the low dust temperatures coupled with energetic processing by X-ray/cosmic ray radiation in the GC. 

%This is a simple template for authors to write new MNRAS papers. The abstract should briefly describe the aims, methods, and main results of the paper. It should be a single paragraph not more than 250 words (200 words for Letters). No references should appear in the abstract.
\end{abstract}

\begin{keywords}
Galaxy: centre -- ISM: molecules -- ISM: abundances --ISM: clouds
\end{keywords}

%%%%%%%%%%%%%%%%%%%%%%%%%%%%%%%%%%%%%%%%%%%%%%%%%%

%%%%%%%%%%%%%%%%% INTRODUCTION %%%%%%%%%%%%%%%%%%%

\section{Introduction}
In the past decades, interstellar molecules with increasing complexity have received significant attention due to their potential prebiotic relevance to the origin of life. These molecules can also be used as tools to constrain the physical properties of the parental environment \citep{calcutt_high-resolution_2014}. Up to date, nearly 200 molecules have been detected in the interstellar or circumstellar medium whilst about one third of them are considered to be complex organic molecules (COMs)\footnote{https://www.astro.uni-koeln.de/cdms/molecules}. In astrochemistry, COMs are generally referred to carbon containing molecules with 6 or more atoms \citep{herbst_complex_2009}. Across various interstellar environments, COMs are considered to be mainly associated with hot cores \citep[e.g. Sgr B2N;][]{hollis_interstellar_2000,hollis_interstellar_2002,hollis_detection_2006,belloche_increased_2009,belloche_complex_2013,belloche_exploring_2016,halfen_formation_2011,halfen_interstellar_2015,rivilla_formation_2017} and hot corinos \citep[e.g. IRAS 16293-2422;][]{van_dishoeck_molecular_1995,bisschop_interferometric_2008,caux_timasss:_2011,kahane_detection_2013,jorgensen_detection_2012,jorgensen2016,jaber_census_2014,jaber_history_2017,ligterink_alma-pils_2017,lykke_alma-pils_2017,martin-domenech_detection_2017} i.e. the compact hot cocoons associated with early stages of high- and low-mass star formation respectively. Apart from these star-forming regions, COMs are now routinely detected toward cold dark cloud cores \citep[e.g. TMC-1;][]{ohishi_chemical_1998,marcelino_discovery_2009,gratier_new_2016,burkhardt_detection_2017,cordiner_deep_2017}, prestellar cores \citep[e.g. Barnard 1 (B1) and L1544;][]{oberg_cold_2010,bacmann_detection_2012,jimenez-serra_spatial_2016,quenard_detection_2017,vastel_origin_2014,vastel_abundance_2016,Vastel2018}, molecular outflows \citep[e.g. L1157;][]{arce_complex_2008,yamaguchi_detection_2011,codella_methyl_2009,codella_seeds_2017,lefloch_chess_2012,lefloch_l1157-b1_2017,mendoza_molecules_2014,Mendoza2018}, and photo-dominated regions (PDR) \citep{gratier_iram-30_2013,cuadrado_chemistry_2015,cuadrado_complex_2017}. Moreover, a variety of COMs have shown their existences in extreme environments such as our own Galatic Centre molecular clouds \citep{requena-torres_organic_2006,requena-torres_largest_2008,martin_tracing_2008,brunken_interstellar_2010,armijos-abendano_3_2014}, and also toward several extragalatic nuclei \citep{martin_galaxies_2011,Aladro2015,Costaliola2015,Harada2018}. 

Although our knowledge of the chemical complexity in the interstellar medium (ISM) has made significant progress thanks to the current development of new instrumentation as well as laboratory experiments, chemical modelling and spectroscopic data, the mechanisms of COMs formation from simple atoms and molecules are still subject to strong debate. Two main formation routes have been proposed to account for the presence of COMs: i) gas-phase chemistry, in which large COMs are formed via gas phase reactions from precursors such as CH$_3$OH or H$_2$CO, which are formed in icy grains and subsequently released to gas phase from grain surfaces \citep{vasyunin_reactive_2013,vasyunin_formation_2017,balucani_formation_2015} and ii) hydrogenation and/or radical-radical reactions on dust grain surfaces, by which COMs are formed entirely within icy mantles and subsequently released to the gas phase \citep{garrod_formation_2006,garrod_complex_2008,taquet_multilayer_2012}. The detection of numerous COMs and the study of their relative abundances in different environments of the ISM is therefore paramount to understand the emergence of molecular complexity and to constrain the main chemical routes (either in the gas phase or on grain surfaces) involved in their formation. 

In this regard, quiescent giant molecular clouds (QGMCs, hereafter) in the Galactic Centre (GC) offer a unique opportunity to better constrain the formation routes of COMs under extreme conditions. Within the Central Molecular Zone \citep[CMZ, the central 500$\,$pc of the Milky Way,][]{morris_galactic_1996}, selected observation of O-bearing molecules carried out across QGMCs in the GC have revealed that COMs are ubiquitous in the whole region \citep{requena-torres_organic_2006}. In comparison to hot cores in the Galactic disk, the physical conditions of the QGMCs in the GC are very different: QGMCs have high gas kinetic temperatures ranging from $\sim$50$\,$K to $\sim$120$\,$K \citep{Guesten1985,huettemeister_kinetic_1993,Rodriguez-Fernandez2001,ginsburg_dense_2016,Krieger2017}, low dust temperatures of $\leq$30$\,$K \citep{Rodriguez-Fernandez2004}, and relatively low H$_2$ gas densities \citep[$\sim$10$^4$$\,$cm$^{-3}$;][]{rodriguez-fernandez_non-equilibrium_2000}. Due to the low H$_2$ densities, the excitation temperatures of the observed COMs are also low \citep[$\sim$10-20$\,$K;][]{requena-torres_organic_2006}, leading to the sub-thermal excitation of the molecules. This is in contrast with the thermal coupling between dust and gas in the Galactic disk hot cores, where the excitation temperature is close to the kinetic temperature.

In order to study the origin of COMs in the GC, \citet{requena-torres_organic_2006} carried out a COMs survey toward several QGMCs in the CMZ and established that these clouds likely represent the largest repository of O-bearing COMs in the Galaxy. Particularly, G+0.693-0.027 (hereafter G+0.693) is one of the GC QGMCs with similar levels of chemical richness and diversity as that found in the prolific GC hot core SgrB2(N) \citep[see e.g.][]{belloche_complex_2013}. The low excitation temperatures in G+0.693 helps in the spectroscopic identification of large COMs since the emission peak in their spectra shifts toward longer wavelengths ($\geq$1mm), avoiding the confusion from the emission of rotational transitions arising from lighter molecular species. \citet{requena-torres_largest_2008} presented the first detection of O-bearing COMs as large as that observed in the star-forming cluster SgrB2(N), such as aldehydes (glycoaldehyde CH$_2$OHCHO, propynal HC$_2$CHO, propenal CH$_2$CHCHO, and propanal CH$_3$CH$_2$CHO) and alcohols (ethylene glycol HOCH$_2$CH$_2$OH and ethylene oxide c-C$_2$H$_4$O) with large column densities. Moreover, more exotic species such as phosphorus bearing species (PN and PO), have also been recently detected towards this source \citep{Rivilla2018}. However, the full inventory of N-bearing COMs in this cloud remained to be explored until now. 

Following the work of \citet{requena-torres_largest_2008} on O-bearing molecules, 
we present the chemical inventory of N-bearing COMs observed toward the quiescent GMC G+0.693 in the CMZ. In Section 2, we report the observations of a GBT 1cm survey (discontinuous frequency covered range between 13-26$\,$GHz) and of IRAM 30m 3 mm, 2 mm, and 1 mm surveys carried out toward this source. We provide not only a rather complete census of the N-bearing species detected but also a comparison of their relative abundances. We also compare our results to the abundances measured across various astrophysical environments ranging from hot cores, to hot corinos, molecular outflows, dark cloud cores, and also several extragalactic sources to understand the possible origin of the COM richness in G+0.693. Finally, we discuss the implications of this study on the synthesis of COMs regarding models based on gas-phase chemistry and reactions on dust grain surfaces.

%%%%%%%%%%%%%%%%%%%%%%%%%%%%%%%%%%%%%%%%%%%%%%%%%%

%%%%%%%%%%% OBSERVATIONS AND RESULTS %%%%%%%%%%%%%

%%%%%%%%%%%%%%%%%%%%%%%%%%%%%%%%%%%%%%%%%%
%%%%%%%%%%% TABLE-OBSERVATIONS %%%%%%%%%%%
%%%%%%%%%%%%%%%%%%%%%%%%%%%%%%%%%%%%%%%%%%

\begin{table*}
\centering
\caption{An overview of IRAM 30 m and GBT observations toward G+0.693 used in this work.}
\begin{tabular}{l c c c c}
\hline
\hline
Date & Frequency coverage & Spectral resolution & Beam size & Telescope \\
 & (GHz) & (km s$^{-1}$) & ($^{\prime\prime}$) & \\
\hline
July 2009-Oct. 2009 & 12-15, 18-26$^a$ & 2.2-8.6 & 29-55 & GBT  \\
July 2009-Sept. 2011 & 80-116 & 5.2-7.5 & 22-29 & IRAM 30m \\
Aug. 2003-Sept. 2005 & 128-176 & 6.9-9.4 & 14-19 & IRAM 30m  \\
July 2004-Dec. 2004 & 240-272 & 2.0 & 9-10 & IRAM 30m  \\
\hline
\hline
\end{tabular}
\label{tab:observations}

\flushleft$^a$ The given frequency ranges are discontinuous, the actual frequencies covered are as following: 12.7-15.5$\,$GHz, 18.12-18.32$\,$GHz, 18.55-18.75$\,$GHz, 18.85-19.06$\,$GHz, 19.16-19.36$\,$GHz, 19.52-19.72$\,$GHz, 20.13-20.38$\,$GHz, 21.36-21.54$\,$GHz, 22.04-22.24$\,$GHz, 23.3-24.45$\,$GHz, 24.84-25-02$\,$GHz, and 25.98-26.18$\,$GHz. 
\end{table*}

\section{Observations}
The spectral line surveys were performed with the IRAM 30m telescope at Pico Veleta (Spain) and the NRAO\footnote{The National Radio Astronomy Observatories is a facility of the National Science Foundation, operated under a cooperative agreement by Associated Universities, Inc.} 100$\,$m Robert C. Byrd Green Bank telescope (GBT) in West Virginia (USA). An overview of the observations toward the QGMC G+0.693 used in this work is presented in Table \ref{tab:observations}. The position switching mode was used in all the observations. The reference position was $\alpha$(J2000.0)= 17$^h$ 46$^m$ 23.01$^s$ and $\delta$(J2000.0)= -28$^{\circ}$ 16$^{\prime}$ 37$^{\prime\prime}$, obtained by \citet{Bally1987}. The coordinates of the quiescent molecular cloud G+0.693-0.03 are $\alpha$(J2000.0)= 17$^h$ 47$^m$ 22$^s$ and $\delta$(J2000.0)= -28$^{\circ}$ 21$^{\prime}$ 27$^{\prime\prime}$. The line intensity of our spectra is given in TA* as the molecular emission toward G+0.693 is extended over the beam \citep{requena-torres_organic_2006,martin_tracing_2008,Rivilla2018}.

% IRAM 30M OBSERVATION %

\subsection{IRAM 30m single-dish observations}
The observations were carried out in multiple sessions between August 2003 and September 2011. In the 2003$-$2005 period, our observations covered the 1 mm (240-272 GHZ) and 2 mm windows (128-176 GHz) with the SIS C and D receivers using the 4$\,$MHz filterbanks (4$\,$MHz spectral resolution). Between 2009 and 2011 the broad-band Eight MIxer Receiver (EMIR) were used at 3mm (80-116$\,$GHz) with the Wideband Line Multiple Autocorrelator (WILMA) which provided a spectral resolution of 2$\,$MHz. The equivalent velocity resolutions in km$\,$s$^{-1}$ are summarised in Table \ref{tab:observations}, and they are high enough to resolve the typical linewidth of the emission measured toward GC QGMCs ($\sim$20$\,$km$\,$s$^{-1}$). The half-power beam widths (HPBW) of the telescopes are in the range of 31$"$-9$"$. Typical system temperatures, $T_{\rm sys},$ ranged between 120-165$\,$K, 180-390$\,$K, and 220-680$\,$K at 3mm, 2mm and 1mm respectively. 

% GBT OBSERVATION %
\subsection{Green Bank Telescope (GBT) observations}
The observations were performed in July-October 2009 (project code GBT09B-015). The Ku-band receiver was connected to the spectrometer, providing four 200$\,$MHz spectral windows with a spectral resolution of 195$\,$kHz (equivalent to a velocity resolution of 2.2-8.6$\,$kms$^{-1}$). Two polarizations were considered during the observations. We calibrated the spectra using a noise tube, providing line intensities affected by 20$\%$ uncertainties.  

%%%%%%%%%%%%%%%%%%%%%%%%%%%%%%%%%%%%%%%%%%%%%%
% ANALYSIS %
\section{Analysis and Results}
\subsection{Line identification and molecular column densities}
  For line identification and analysis, our spectra were exported from the GILDAS/CLASS software package\footnote{http://www.iram.fr/IRAMFR/GILDAS.} to the MADCUBA package\footnote{Madrid Data Cube Analysis on ImageJ is a software developed in the Centre of Astrobiology (Madrid, INTA-CSIC) to visualise and analyse single spectra and datacubes \citep[][Mart\'in et al., in prep.]{rivilla_first_2016}}. MADCUBA contains molecular databases such as JPL\footnote{http://spec.jpl.nasa.gov/}\citep{pickett_submillimeter_1998} and CDMS\footnote{http://www.astro.uni-koeln.de/cdms}\citep{muller_cologne_2001,muller_cologne_2005,endres_cologne_2016} that provide the frequencies and the spectroscopic information of different species. With the identified transitions, the Spectral Line Identification and Modelling (SLIM) package implemented within MADCUBA was used to produce synthetic spectra by assuming Local Thermodynamical Equilibrium (LTE) conditions and considering the effect of line opacity. These synthetic spectra were compared to the observed spectra where the parameters of total molecular column density, $N_{\rm tot}$, excitation temperature, $T_{\rm ex}$, peak velocity, $V_{\rm LSR}$, and linewidth, $\Delta \nu$ of the emission are fit. Then, the MADCUBA-AUTOFIT tool was able to provide the best non-linear least-squared fit using the Levenberg-Marquardt algorithm. A representative sample of the observed and fitted line profiles for several molecular transitions observed in G+0.693 are shown in Figure \ref{fig-line-profiles}, and the complete set of spectra for detected molecular lines are presented in Appendix B. The physical parameters together with their associated uncertainties derived from the MADCUBA AUTOFIT are listed in Table \ref{tab:gaussian-fit-parameters}. Note that for some molecules such as ortho-H$_2$CCN and HC$_7$N, the AUTOFIT algorithm would only converge if the value of $V_{\rm LSR}$ and/or $\Delta \nu$ were fixed (see the corresponding values in Table \ref{tab:gaussian-fit-parameters}). Since these parameters are fixed to a certain value by the user, no uncertainties are associated with $V_{\rm LSR}$ and/or $\Delta \nu$ within  MADCUBA during the line fitting process. In addition, $T_{\rm ex}$ of a given species can only be constrained if multiple rotational transitions are available; otherwise $T_{\rm ex}$ needs to be fixed as well. 
  
  As shown in Table \ref{tab:gaussian-fit-parameters}, the $N_{\rm tot}$ of some molecules are attributed to different type of transitions: a- or b-type transitions for formamide (NH$_2$CHO), ortho/para transitions for cyanomethyl radical (H$_2$CCN) and cyanamide (NH$_2$CN), and K-ladders/K$_a$ ladders present in methyl cyanide (CH$_3$CN) and isocyanic acid (HNCO) respectively. In particular, rotational diagram analysis is used to estimate the excitation temperature and column density of CH$_3$CN due to the presence of several series of K-ladder transitions (more details in Section 3.3). The main spectroscopic information for these molecules are summarised in Appendix A1. 
  
  In total, we have identified 19 N-bearing molecules excluding their isotopologues toward G+0.693. Among them, there are 17 clear detections with peak intensities above 3$\sigma$, and 2 tentative detections (where only one or two transitions showed peak intensities higher than 3$\sigma$, see Figure \ref{fig-tentative-detection}). For the non-detections, we estimated  3$\sigma$ integrated line intensity noise levels as 3 $\times$ rms $\times \sqrt{\Delta \nu \times \delta v}$, where rms (in K) is estimated with MADCUBA, $\Delta \nu$ is the assumed line widths of the transition (~20$\,$km$\,$s$^{-1}$) and $\delta v$ is the velocity spectral resolution in km$\,$s$^{-1}$. The 3$\sigma$ column densities and corresponding relative abundances with respect to H$_2$ assuming $T_{\rm ex}$=15-20$\,$K are listed in Table \ref{tab:non-detections}.

%%%%%%%%%%%%%%%%%%%%%%%%%%%%%%%%%%%%%%%%%%
%%%%%% TABLE-GAUSSIAN-FIT-PARAMETERS %%%%%
%%%%%%%%%%%%%%%%%%%%%%%%%%%%%%%%%%%%%%%%%%
\begin{table}
 \centering
 \caption{Physical parameters of the N-bearing molecules derived from the MADCUBA-LTE analysis of molecular spectra for G+0.693}
\begin{adjustbox}{width=0.5\textwidth}
\begin{tabular}{ccccc}
\hline
\hline
 Formula & $T_{\rm ex}$ & $V_{\rm LSR}$ & $\Delta \nu$ & $N_{\rm tot}$ \\
  & (K) & (km s$^{-1}$) & (km s$^{-1}$) & ($\times$10$^{13}$cm$^{-2}$) \\
\hline
\bf{NO} & 4.1$\pm$0.4 & 68$\pm$1 & 21$\pm$1 & 3600$\pm$200 \\
\hline
HNCO, K$_a$=0 & 17$\pm$1 & 67$\pm$1 & 23$\pm$1 & 320$\pm$11 \\
HNCO, K$_a$=1 & 29$\pm$2 & 68$\pm$1 & 22$\pm$1 & 16$\pm$1 \\
\bf{HNCO}$^a$ & - & - & - & 336$\pm$12 \\
\hline
\bf{HOCN} & 8.1$\pm$0.3 & 69$\pm$1	& 20$\pm$1 & 1.9$\pm$0.1 \\
\bf{C$_3$N}	& 7.3$\pm$1.5 & 68$\pm$1 & 18$\pm$2 & 4.3$\pm$0.2 \\
\bf{CH$_2$NH} & 9.7$\pm$0.4 & 69$\pm$1 & 25$\pm$1 & 54$\pm$3 \\
\hline
H$^{13}$CCCN & 12$\pm$2 & 68$\pm$1 & 22$\pm$1 & 3.4$\pm$0.6 \\
\bf{HC$_3$N}$^b$ & - & - & - & 71$\pm$13 \\
\hline
\bf{HCCNC} & 7.0$\pm$2.1 & 71$\pm$1 & 19$\pm$2 & 2.3$\pm$0.7 \\
\hline
o-H$_2$CCN$^c$ & 12$\pm$1 & 70 & 30 & 5.4$\pm$0.3 \\
p-H$_2$CCN & 9.1$\pm$0.6 & 70$\pm$1 & 21$\pm$1 & 17$\pm$2 \\
\bf{(o+p)-H$_2$CCN}$^d$ & - & - & - & 22$\pm$2 \\
\hline
o-NH$_2$CN & 6.3$\pm$0.3 & 66$\pm$1 & 24$\pm$1 & 3.8$\pm$0.2 \\
p-NH$_2$CN & 6.8$\pm$0.2 & 67$\pm$1 & 24$\pm$1 & 27$\pm$2 \\
\bf{(o+p)-NH$_2$CN}$^d$  & - & - & - & 31$\pm$2 \\
\hline
CH$_3$CN(J=5-4) & 140$\pm$26 & 69$\pm$1 & 27$\pm$1 & 100$\pm$22 \\
CH$_3$CN(J=6-5) & 100$\pm$8 & 69$\pm$1 & 25$\pm$1 & 46$\pm$5 \\
CH$_3$CN(J=7-6) & 78$\pm$6 & 69$\pm$1 & 26$\pm$1 & 27$\pm$3 \\	
CH$_3$CN(J=8-7) & 73$\pm$6 & 68$\pm$1	& 26$\pm$1 & 15$\pm$1 \\
CH$_3$CN(J=9-8) & 85$\pm$11 & 68$\pm$1	& 25$\pm$1 & 13$\pm$2 \\	
CH$_3$CN(J=14-13) & 83$\pm$9 & 70$\pm$1 & 23$\pm$1 & 0.9$\pm$0.1 \\
CH$_3$CN(J=15-14)$^c$ & 130 & 71$\pm$1 & 16$\pm$2	& 0.40$\pm$0.04	\\
\bf{CH$_3$CN}$^e$ & 14 & 71 & 21 & 11.5$\pm$0.3 \\
\hline
a-NH$_2$CHO & 9.8$\pm$0.6 & 71$\pm$1 & 23$\pm$1 & 5.5$\pm$0.3 \\
b-NH$_2$CHO & 3.7$\pm$0.7 & 68$\pm$1 & 23$\pm$2 & 57$\pm$14 \\
\bf{NH$_2$CHO}$^f$ & & & & 63$\pm$14 \\
\hline
\bf{HC$_5$N} & 16.2$\pm$0.2 & 67$\pm$1 & 22$\pm$1 & 26$\pm$1 \\
\bf{C$_2$H$_3$CN} & 10.8$\pm$1.1 & 68$\pm$1 & 22$\pm$2 & 9$\pm$1 \\
\bf{CH$_3$NH$_2$} & 16.2$\pm$0.8 & 67$\pm$1 & 20$\pm$1 & 30$\pm$9\\
\hline
\bf{CH$_3$NCO} & 7.9$\pm$0.4 & 67$\pm$1 & 23$\pm$2 & 6.6$\pm$0.4\\
\bf{C$_2$H$_5$CN} & 17.7$\pm$1.6	& 69$\pm$1 & 24$\pm$2 & 4.1$\pm$0.4 \\
\bf{HC$_7$N} & 7$\pm$1 & 66 & 26$\pm$2 & 1.5$\pm$0.3 \\
\hline
\multicolumn{5}{c}{Tentative Detections} \\
\hline
HNCCC & 5.3$\pm$0.5 & 68$\pm$1 & 14$\pm$3 & $\leq$0.2$\pm$0.1 \\
CH$_3$C$_3$N$^c$ & 15 & 68 & 20 & $\leq$4.5$\pm$0.3	\\
\hline
\hline
\end{tabular}
\end{adjustbox}
\label{tab:gaussian-fit-parameters}
\begin{tablenotes}
\item $^{(a)}$ $N_{\rm tot}$ calculated by the sum of K$_a$=0 and K$_a$=1 rotational ladder. $^{(b)}$ $N_{\rm tot}$ was derived from its isotopologue H$^{13}$CCCN due to the emission lines of HC$_3$N are optically thick. We adapted the $^{12}$C/$^{13}$C $\sim$ 21 from \citep{armijos-abendano_3_2014} in G+0.693. $^{(c)}$ Values fixed. $^{(d)}$ N$_{tot}$ calculated by the sum of the ortho and para species. $^{(e)}$ $T_{\rm ex}$ and $N_{\rm tot}$ derived from rotational diagram analysis (see Figure \ref{fig-rotational-diagram}). $^{(f)}$ $N_{\rm tot}$ calculated by the sum of a-type and b-type transitions.
\end{tablenotes}
\end{table}

%%%%%%%%%%%%%%%%%%%%%%%%%%%%%%%%%%%%%%%%%%
%%%%%%%%% TABLE-UNDETECTED-COMS %%%%%%%%%%
%%%%%%%%%%%%%%%%%%%%%%%%%%%%%%%%%%%%%%%%%%
\begin{table}
 \centering
 \caption{Upper limits for undetected N-bearing species}
\begin{tabular}{lcc}
\hline
\hline
 Formula & $N_{\rm tot}$ & Abundance	\\
 & ($\times$10$^{13}$cm$^{-2}$) & ($\times$10$^{-11}$) \\
\hline
C$_2$N & $\leq$0.3 & $\leq$1.9 \\
NCO	& $\leq$2 & $\leq$11.7 \\
H$_2$CN  & $\leq$0.3 & $\leq$1.9 \\
HCCN & $\leq$1.1 & $\leq$8.2 \\
HCNO & $\leq$0.3 & $\leq$2.5 \\
HONC & $\leq$0.1 & $\leq$0.7\\
C$_4$N	& $\leq$0.4 & $\leq$2.9\\
HCOCN & $\leq$6 & $\leq$44.7 \\
NH$_2$OH  & $\leq$2 & $\leq$14.1 \\
l-HC$_4$N	& $\leq$1 & $\leq$7.4 \\
C$_5$N	& $\leq$0.6 & $\leq$4.7 \\
CH$_3$NC & $\leq$0.3 & $\leq$2.3 \\
(NH$_2$)$_2$CO & $\leq$1 & $\leq$6.8 \\
C$_2$H$_3$NC & $\leq$0.3 & $\leq$2.2 \\
CH$_2$(CN)$_2$ & $\leq$2.5 & $\leq$18.6 \\
CH$_3$OCN	& $\leq$0.6 & $\leq$4.7 \\
CH$_3$CNO	& $\leq$0.1 & $\leq$0.8 \\
NCCONH$_2$	& $\leq$0.3 & $\leq$2.5 \\
l-HC$_6$N	& $\leq$0.3 & $\leq$2.3 \\
CH$_3$CCNC	& $\leq$0.3 & $\leq$2.3 \\
H$_2$CCCHCN	& $\leq$2 & $\leq$14.8 \\
H$_2$NCH$_2$CN	& $\leq$0.6 & $\leq$4.7 \\
C$_2$H$_3$NH$_2$ & $\leq$4 & $\leq$29.5\\
c-C$_2$H$_4$NH	& $\leq$4 & $\leq$29.5 \\
CH$_3$C$_5$N	& $\leq$0.1 & $\leq$0.9 \\
HC$_9$N	& $\leq$0.4 & $\leq$2.6 \\
NH$_2$CH$_2$CH$_2$OH & $\leq$1 & $\leq$8.9 \\
i-C$_3$H$_7$CN	& $\leq$0.6 & $\leq$4.7 \\
n-C$_3$H$_7$CN & $\leq$1.1 & $\leq$8.2 \\
HC$_{11}$N	& $\leq$0.2 & $\leq$1.9 \\
HC$_{13}$N	& $\leq$1 & $\leq$1.3\\
\hline
\hline
\end{tabular}
\label{tab:non-detections}
\end{table}

%%%%%%%%%%%%%%%%%%%%%%%%%%%%%%%%%%%%%%%%%%
%%%%%%%% FIGURE-EXAMPLE SPECTRA %%%%%%%%%%
%%%%%%%%%%%%%%%%%%%%%%%%%%%%%%%%%%%%%%%%%%
\begin{figure*}
\includegraphics[width = \textwidth]{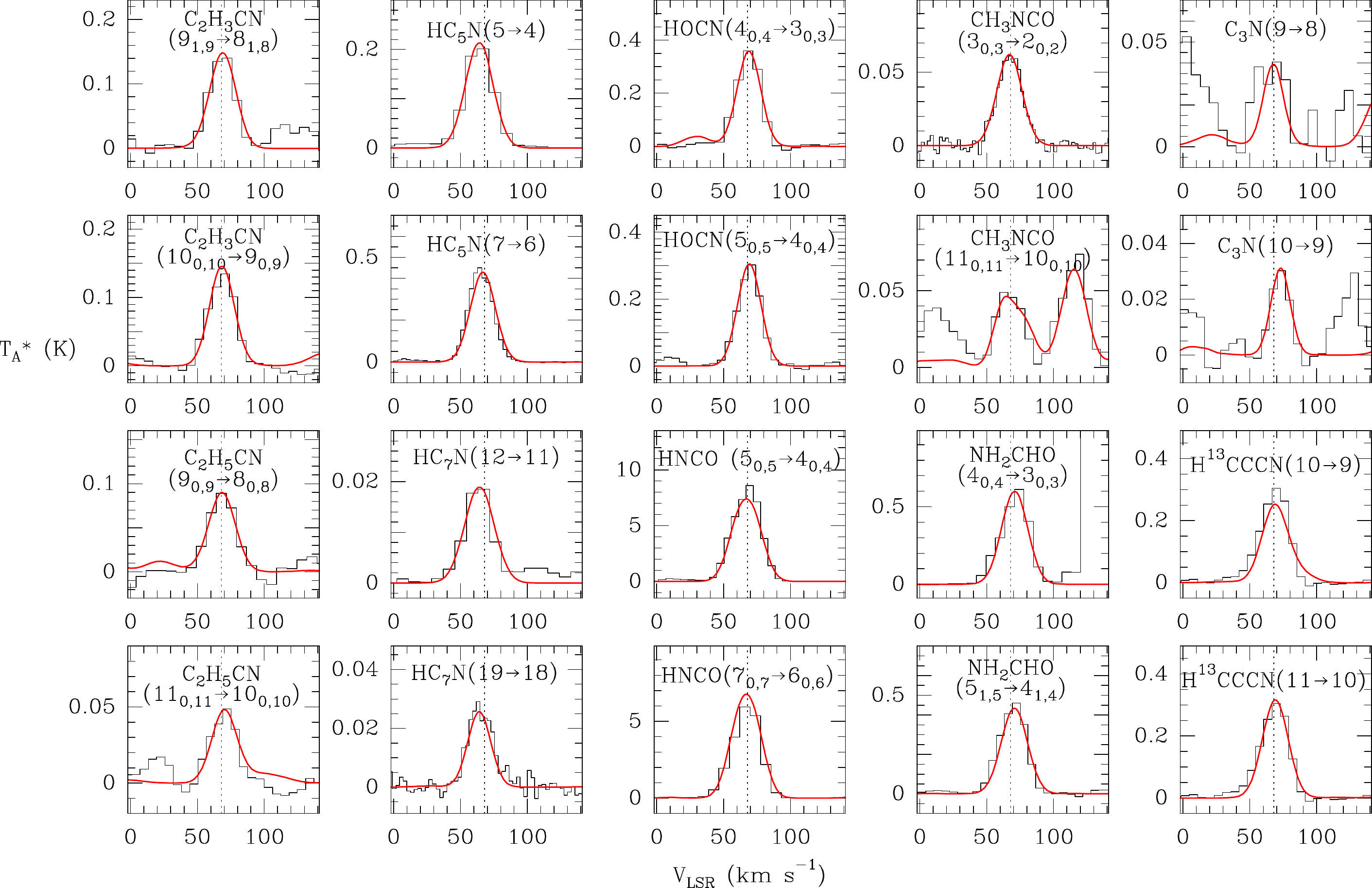}
\vspace{-2mm}
 \caption{Sample line profiles of a number of detected N-bearing molecules in G+0.693. The LTE best fits from MADCUBA are shown in red lines.}
    \label{fig-line-profiles}
\end{figure*}

%%%%%%%%%%%%%%%%%%%%%%%%%%%%%%%%%%%%%%%%%%%%%%%%%%

%%%%%%%%%%%%%%%%%%%%%%%%%%%%%%%%%%%%%%%%%%
%%%%%%%% FIGURE-TENTATIVE-SPECTRA %%%%%%%%
%%%%%%%%%%%%%%%%%%%%%%%%%%%%%%%%%%%%%%%%%%
\begin{figure*}
\includegraphics[width = 0.5\textwidth]{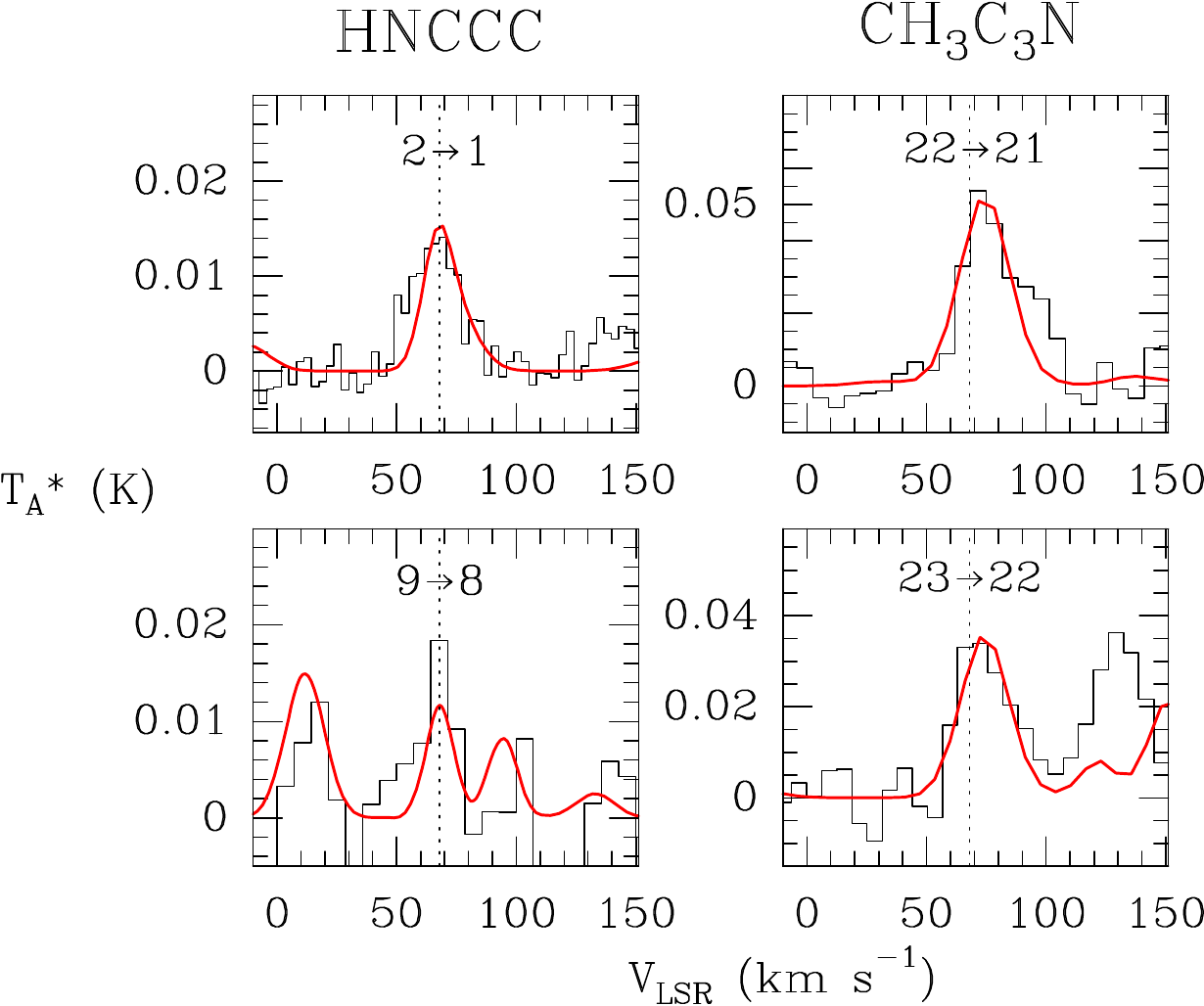}
\vspace{-1mm}
 \caption{Line profiles are tentative detections of molecules HNCCC and CH$_3$C$_3$N. The LTE best fits from MADCUBA are shown in red lines.}
    \label{fig-tentative-detection}
\end{figure*}

%%%%%%%%%%%%%%%%%%%%%%%%%%%%%%%%%%%%%%%%%%%%%%%%%%
 \subsection{Rotational diagram analysis for CH\texorpdfstring{$_3$}CN}
CH$_3$CN is a symmetric rotor in which its rotational energy levels (or J transitions) are further divided into successive K components. Consequently, spectral lines due to different K values are shifted in frequency with respect to each other, giving rise to a so-called 'K-ladder' spectrum. Due to the fact that radiative transitions are forbidden between K-ladders, the level populations are expected to be thermalised and therefore determined by collisional excitation within K-ladders. Then the $T_{\rm ex}$ of the K ladder levels for the J+1$\rightarrow$J transition of CH$_3$CN provides a measurement of the $T_{\rm kin}$ of the source. In our survey, the K-ladders of J=5$\rightarrow$4, J=6$\rightarrow$5, J=7$\rightarrow$6, J=8$\rightarrow$7, J=9$\rightarrow$8, J=14$\rightarrow$13, and J=15$\rightarrow$14 are observed for CH$_3$CN in G+0.693. The individual $T_{\rm ex}$ obtained from these K ladders lie between 73-140$\,K$ which indicates that the $T_{\rm kin}$ can be up to 140$\,K$. These values are consistent with the gas kinetic temperatures $T_{\rm kin}$$\sim$50-120$\,$K previously reported by \citet{Guesten1985,huettemeister_kinetic_1993,Krieger2017} using another symmetric top molecule, ammonia (NH$_3$).

\begin{figure}
\includegraphics[width =\columnwidth]{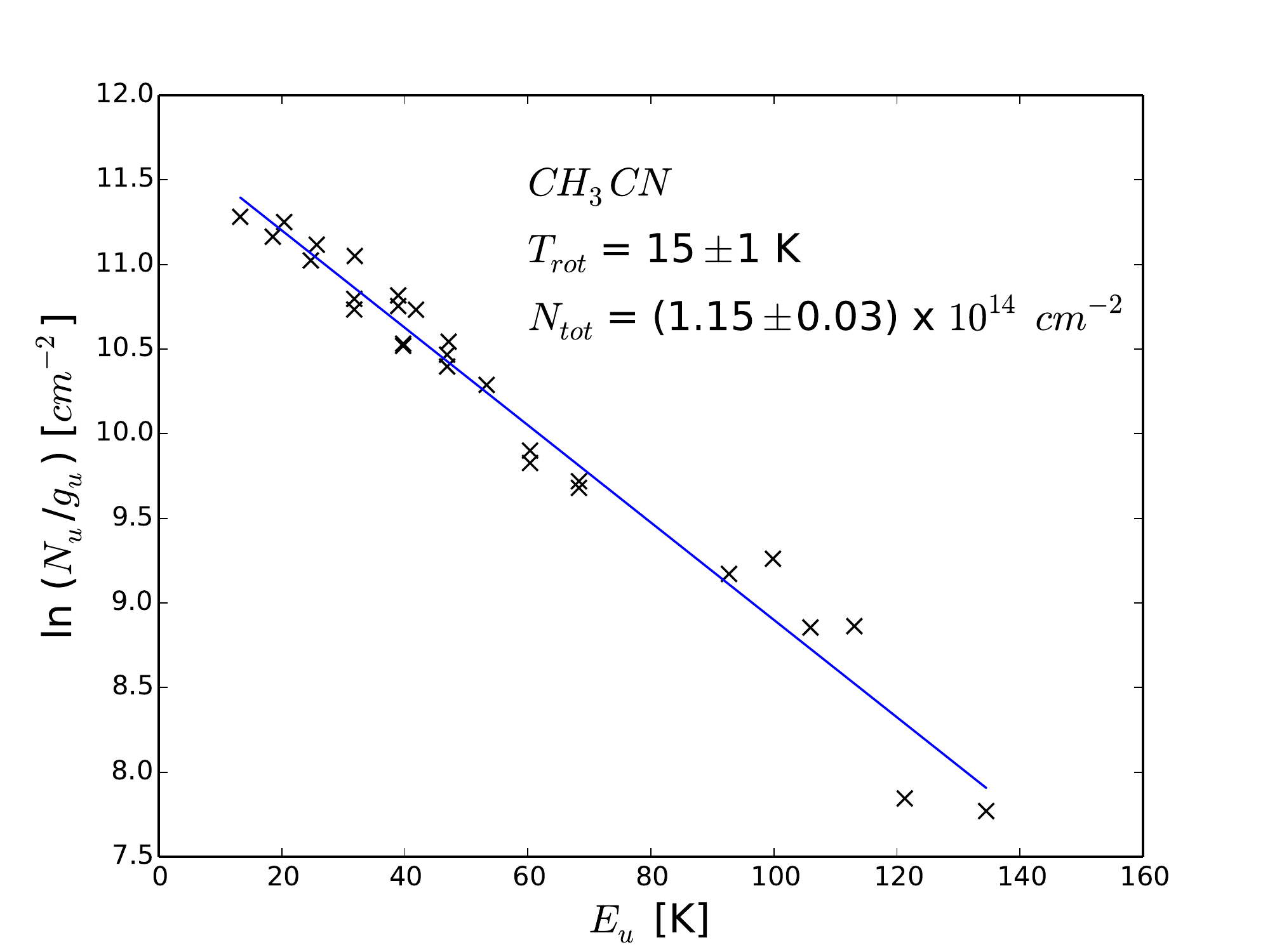}
\vspace{-5mm}
 \caption{CH$_3$CN rotational diagram constructed using the K=0, 1, and 2 transitions for all detected J+1$\rightarrow$J transitions. The blue straight line indicates the best linear regression fit to the data points.}
    \label{fig-rotational-diagram}
\end{figure}

In addition, a rotational diagram was constructed for CH$_3$CN to derive the overall $N_{\rm tot}$ and $T_{\rm rot}$ (see Figure {\ref{fig-rotational-diagram}}). A linear regression line was fitted to the lines with the lowest K values i.e. K=0, 1, and 2 for all J+1$\rightarrow$J transitions, yielding $T_{\rm rot}$=15$\pm$1$\,$K. By fixing $T_{\rm rot}$=15$\,$K, $V_{\rm LSR}$=71$\,$km s$^{-1}$, and $\Delta \nu$=21$\,$km s$^{-1}$ in MADCUBA, the AUTOFIT algorithm converged to give a $N_{\rm tot}$(CH$_3$CN) value of (1.15$\pm$0.03)$\times$10$^{14}$$\,$cm$^{-2}$. 

\subsection{Excitation temperature vs gas kinetic temperature}  
The $T_{\rm ex}$ derived from the source using different molecules range between 9-30$\,K$. Note that the $T_{\rm rot}$=15$\pm$1$\,$K extracted from the rotational diagram of CH$_3$CN, also falls in the same range. This range of $T_{\rm ex}$ is consistent with those found in \citet{requena-torres_organic_2006,requena-torres_largest_2008}, and significantly lower that the derived kinetic temperatures of the gas ($T_{\rm kin}$=73-140 K; see Section 3.2). This implies that the molecular emission in this cloud is sub-thermally excited due to its relatively low H$_2$ gas densities \citep[see][]{rodriguez-fernandez_non-equilibrium_2000}. 

Overall, from the detected N-bearing molecules, we derive a range of $T_{\rm ex}$ and $T_{\rm kin}$ that are consistent with previous studies. They both strengthen the fact that G+0.693 indeed has a much lower $T_{\rm ex}$ compared to its $T_{\rm kin}$, a signature of the sub-thermal excitation of molecules within the cloud.
\\

%%%%%%%%%%%%%%%%%%%%%%%%%%%%%%%%%%%%%%%%%%
%%%%%%% FIGURE-ROTATIONAL DIAGRAM %%%%%%%%
%%%%%%%%%%%%%%%%%%%%%%%%%%%%%%%%%%%%%%%%%%

\subsection{Molecular abundances relative to \texorpdfstring{H$_2$}, and parent molecule of each functional group}
Since we know that the emission is extended, the relative abundances can be estimated by dividing the measured molecular column densities by the H$_2$ column density ($N_{\rm H_2}$). In this study, we adopt the H$_2$ column density $N_{\rm H_2}$ = 1.35 $\times$ 10$^{23}$ cm$^{-2}$ derived by \citet{martin_tracing_2008} from C$^{18}$O. In our calculations, we assume that all molecules present a similar spatial distribution to C$^{18}$O when determining the observed abundances (i.e. all molecules arise from the same region). For molecules with clear detections and tentative detections, high abundances relative to H$_2$ are observed toward G+0.693, ranging from 10$^{-11}$ to 10$^{-8}$. Uncertainties in the relative abundances can be estimated by propagating errors in the molecular column densities provided by the MADCUBA AUTOFIT algorithm. A summary of the calculated abundances relative to H$_2$ is given in the second column of Tables \ref{tab:abundance-CN}, \ref{tab:abundance-NH}, and \ref{tab:abundance-OCN}. In order to explore the possible differences in chemistry, in Tables \ref{tab:abundance-CN}, \ref{tab:abundance-NH}, and \ref{tab:abundance-OCN}, we also report the molecular abundance ratios between each molecule and their corresponding "parent" species within each functional group: CH$_3$CN for nitriles (-CN), CH$_3$NH$_2$ for amines (-NH), and HNCO for cyanates (-NCO). Note that we refer here to parent species as the most abundant molecules of each functional group.

%%%%%%%%%%%%%%%%%%%%%%%%%%%%%%%%%%%%%%%%%%
%%%%%%%% TABLE-ABUNDANCE-CN-GROUP %%%%%%%%
%%%%%%%%%%%%%%%%%%%%%%%%%%%%%%%%%%%%%%%%%%

 \begin{table}
  \centering
  \caption{Abundances ratios with respect to H$_2$ and CH$_3$CN.}
  \begin{tabular}{lcc}
  \hline
   Molecule & X(H$_2$) & X(CH$_3$CN)\\
    & ($\times$10$^{-10}$) & \\
  \hline
  \hline
   CH$_3$CN & 9 & 1 \\
   H$_2$CCN & 16 & 2 \\
   C$_3$N & 3 & 0.4 \\
   HCCCN & 54 & 6 \\
   HCCNC & 2 & 0.2 \\
   HNCCC & $\leq$0.1 & $\leq$0.02\\
   C$_2$H$_3$CN & 7 & 0.8 \\
   C$_2$H$_5$CN & 3 & 0.4 \\
   CH$_3$C$_3$N & $\leq$3 & $\leq$0.4 \\
   HC$_5$N & 19 & 2.21 \\
   HC$_7$N & 1 & 0.1 \\
   *NH$_2$CN & 23 & 3 \\
  \hline
  \hline
  \end{tabular}
  \label{tab:abundance-CN}
  \begin{tablenotes}
  \item (*) denote molecule that potentially contains two functional groups i.e. NH$_2$CN contains both NH- and CN- group. 
  \end{tablenotes}
\end{table}

%%%%%%%%%%%%%%%%%%%%%%%%%%%%%%%%%%%%%%%%%%
%%%%%%% TABLE-ABUNDANCE-NH-GROUP %%%%%%%%%
%%%%%%%%%%%%%%%%%%%%%%%%%%%%%%%%%%%%%%%%%%

\begin{table}
  \centering
  \caption{Abundances ratios with respect to H$_2$ and CH$_3$NH$_2$.}
  \begin{tabular}{lcc}
  \hline
   Molecule & X(H$_2$) & X(CH$_3$NH$_2$) \\
    & ($\times$10$^{-10}$) & \\ \hline \hline
  %\hline
  %\hline
  CH$_3$NH$_2$ & 221 & 1 \\
  CH$_2$NH & 43 & 0.2 \\
  %H$_2$CCNH & $\leq$25 & $\leq$11 \\
  *NH$_2$CN & 23 & 0.1 \\
  *NH$_2$CHO & 47 & 0.2 \\
  \hline
  \end{tabular}
  \label{tab:abundance-NH}
  \begin{tablenotes}
  \item (*) denote molecules that potentially contain two functional groups i.e. NH$_2$CN contains both NH- and CN-group and NH$_2$CHO contains both NH- and NCO-group.
  \end{tablenotes}
\end{table}

%%%%%%%%%%%%%%%%%%%%%%%%%%%%%%%%%%%%%%%%%%
%%%%%%% TABLE-ABUNDANCE-OCN-GROUP %%%%%%%%%
%%%%%%%%%%%%%%%%%%%%%%%%%%%%%%%%%%%%%%%%%%

\begin{table}
  \centering
  \caption{Abundances ratios with respect to H$_2$ and HNCO.}
  \begin{tabular}{lcc}
  \hline
   Molecule & X(H$_2$) & X(HNCO) \\
    & ($\times$10$^{-10}$) & \\
  \hline
  \hline
   HNCO & 249 & 1 \\
   HOCN & 1 & 0.01 \\
   *NH$_2$CHO & 47 & 0.2 \\
   CH$_3$NCO & 5 & 0.02 \\
 \hline
  \end{tabular}
  \label{tab:abundance-OCN}
  \begin{tablenotes}
  \item (*) denote molecules that potentially contain two functional groups i.e. NH$_2$CHO contains both NH- and NCO-group.
  \end{tablenotes}
\end{table}

%%%%%%%COMPARISON OF ABUNDANCE RATIOS%%%%%%%%%%%%%%%
\section{Discussion}
As for the O-bearing family in \citet{requena-torres_largest_2008}, the full census of the identified N-bearing molecules in G+0.693 are presented in Figures \ref{fig-N-bearing-N} and \ref{fig-N-bearing-O}. The purpose of these diagrams is to organize N-bearing molecules by increasing complexity (increasing number of carbon C, nitrogen N, and oxygen O atoms) so that the possible relation between two molecules, or even from a simple molecule to a very complex one, can be quickly visualised in one diagram. Besides molecules, intermediate species such as radicals and chemically unstable species, which presumably represent the missing links between chemical reactions, are also included in the diagram. Molecules that have not been identified in space either due to lack of spectroscopic data or lack of sensitivity in observations are also indicated. The measured molecular abundances with respect to H$_2$ and the derived upper limits for the non-detections are both given in the diagrams. These diagrams could be used as guidance for future searches of species as well as for spectroscopic laboratory experiments.

\subsection{Comparison of relative abundances across different galactic environments}
It has been proposed that the formation processes of COMs in hot and cold environments largely differ. In hot cores and hot corinos, COMs are believed to form via radical-radical association on the surface of dust grains as radicals become mobile at temperatures $\geq$30$\,$K during the warming-up of the protostellar envelope \citep{garrod_complex_2008}. The heating from the protostar sublimates the ices (and its content in COMs) off dust grains, triggering a complex chemistry in the hot envelope. In contrast to these hot environments, the gas temperature in cold dark cloud cores rarely exceeds 15$\,$K, and therefore the chemistry is dominated by ion-molecule reactions in the gas phase since the gas radical mobility necessary for COM formation on grains does not occur. As a result of low gas temperature, highly reactive species, including ions, radicals, and unsaturated species such as cyanopolyynes HC$_n$N are abundant in cold core sources \citep{ohishi_chemical_1998,smith_rapid_2004}. COMs have also been found in shocked regions, where the gas is enriched by ice sputtering \citep{arce_complex_2008} and/or affected by gas-phase chemistry \citep{codella_seeds_2017}. The origin of the formation of COMs in G+0.693 has been rather elusive due to its peculiar physical conditions compared to other environments. The ejection of COMs from grain mantles in G+0.693 has been proposed to be due to non-dissociative shocks with  velocities $\leq$40$\,$km s$^{-1}$ \citep{requena-torres_organic_2006,requena-torres_largest_2008,martin_tracing_2008}. This seems to be consistent with the typical line widths ($\sim$20$\,$km $^{-1}$) derived from the molecular transitions.

In order to better understand the formation mechanisms of COMs in G+0.693, we compared the derived abundances relative to H$_2$ of the detected N-bearing molecules across different astronomical environments. These sources include the hot cores Sgr B2(N) and Orion KL, the hot corino IRAS16293-2422 (IRAS16293 hereafter), the dark cloud TMC-1 and the shocked region L1157-B1. As illustrated in Figure \ref{fig-fractional-abundance}, the abundances relative to H$_2$ are plotted for detected and tentatively-detected species which can be categorised into 4 groups: 1) cyanopolyynes HC$_n$N, 2) nitriles -CN, 3) amines -NH, and 4) cyanates -NCO. Within each group, species are organised by increasing complexity i.e. increase number of C, O, and N from left to right. For completeness, the abundance ratios between relative molecules such as isomers and "parents/daughters" species are listed in Table \ref{tab: abundance-ratio}.
\\
%In the following subsections, we compare the abundances of N-bearing species across sources considering the different N-bearing molecular families: cyanopolyynes, nitriles, amines and cyanates. 

\subsubsection{Cyanopolyynes HC$_n$N group}
Linear carbon chain radical C$_3$N as well as cyanopolyynes chains, HC$_n$N (n=3,5,7 etc.) ranging from HC$_3$N through HC$_7$N, have been detected toward G+0.693 in this study. Besides the metastable isomer HC$_3$N, the isomers HCCNC and HNCCC have also been detected. The main isomer HC$_3$N seems to be ubiquitous in the ISM as it has been detected in all sources with relatively high abundances ($\geq$10$^{-9}$; Figure 4 and Table 7). Its HC$_3$N/H$_2$ abundance obtained in G+0.693 is a factor of 5 lower than those found toward L1157-B1 and TMC-1 whilst its HC$_5$N/H$_2$ abundance is consistent with that measured in L1157-B1 \citep{Mendoza2018}. As shown in Table \ref{tab: abundance-ratio}, the relative abundance for molecules following the sequence from HC$_3$N to HC$_7$N is reduced by approximately a factor of 10 with increasing molecular size in G+0.693. Although the abundance ratio between HC$_3$N : HC$_5$N ($\sim$1 : 0.3) matches very well between G+0.693 and TMC-1, both of the HC$_3$N : HC$_7$N and HC$_3$N : HC$_9$N ratio differ by an order of magnitude respectively between these two regions. Recently, cyanopolyynes have also been detected in the intermediate mass protocluster OMC2-FIR4 by \citet{fontani_solis_2017}. HC$_3$N : HC$_5$N abundance ratio (between 1 : 0.08 to 1 : 0.25) have been measured toward the the eastern region of FIR4 where strong HC$_5$N emission is found. Through their analysis along with relevant gas-phase chemical model, it has been suggested that HC$_3$N : HC$_5$N $\geq$0.08 obtained in FIR4 is possibly due to an enhanced cosmic-ray ionisation rate of $\zeta$=4$\times$10$^{-14}$s$^{-1}$. Toward G+0.693, the HC$_3$N : HC$_5$N abundance ratio is 1 : 0.3, consistent with the range obtained in FIR4. This might suggest that cosmic rays can also affect the chemistry of cyanopolyynes in G+0.693 \citep[note that the cosmic-ray ionisation rate in the CMZ is constrained to be $\zeta\sim$1-10$\times$10$^{-15}$s$^{-1}$ which is factors 4-40 lower than that in FIR4;][]{yusef-zadeh_interacting_2013,ginsburg_dense_2016}. 

Alternatively, a low HC$_3$N : HC$_5$N abundance ratio could be the result of a cold temperature chemistry as proposed by \citet{jaber_history_2017}. These authors inferred HC$_3$N : HC$_5$N abundance ratios of 1 : $\sim$10 and 1 : $\sim$1 toward the inner hot corino and the outer cold envelope of IRAS16293, respectively. The HC$_3$N : HC$_5$N ratio of the cold envelope is comparable to that of G+0.693. In their models, \citet{jaber_history_2017} propose that an enhanced cosmic ray ionization rate has little effect on the abundance of HC$_3$N, although their predicted abundance of HC$_5$N does not match the observations. Note also that, while the cold envelope of IRAS16293 has a kinetic temperature of T$_{\rm kin}$=20 K, the T$_{\rm kin}$ in G+0.693 is 73-140 K (see Section 3.2). This implies that the cold chemistry characteristic of the envelope of IRAS16293 cannot be applied to G+0.693.

Being the simplest member of the cyanopolyynes, the most important formation pathway of HC$_3$N has been proposed to be the neutral-neutral reaction (mechanism 1 hereafter) between hydrocarbon molecules C$_n$H$_2$ (n=1, 2,3,4..) and the CN radical. For instance, the gas-phase reaction between C$_2$H$_2$ and CN to form HC$_3$N \citep[see equation (1) in][]{burkhardt_detection_2017} has been supported by several studies toward dark clouds, low-mass and high-mass star-forming regions \citep{takano_observations_1998,burkhardt_detection_2017,taniguchi_13c_2016,taniguchi_$^13$c_2017}. Subsequently, larger cyanopolyynes such as HC$_5$N and HC$_7$N would be assumed to be produced from CN to the same extent as HC$_3$N. A recent detection of the isotopologues of HC$_5$N and HC$_7$N in TMC-1 by \citet{burkhardt_detection_2017} has provided evidence that HC$_5$N and HC$_7$N do not necessarily undergo the same formation pathway as suggested for HC$_3$N. Instead, the dominant formation mechanism for these two cyanopolyynes could be the reaction in gas phase of hydrocarbon ions and nitrogen atoms \citep[mechanism 2 hereafter, see equation (3) in][]{burkhardt_detection_2017}. Since the gas kinetic temperature in G+0.693 is high (73 - 140$\,$K), we have investigated how the reaction rates of these two mechanisms vary with temperature to evaluate their efficiency\footnote{Reaction rates extracted from the KIDA database (http://kida.obs.ubordeaux1.fr/).}. For mechanism 1, the reaction rates remain almost constant (they vary by less than a factor of 2) with increasing temperature from 10K to 100K. This  cannot explain a factor of 5 difference in the abundance of HC$_3$N between G+0.693 and TMC-1. Hence we propose that gas-phase reactions involving CN might not be the main formation route for HC$_3$N. For mechanism 2, however, the reaction rates slow down by a factor of $\sim$5 from 10 to 100$\,$K, which would explain why HC$_5$N and HC$_7$N systematically show lower abundances in G+0.693 compared to TMC-1. All these point toward the idea that CN is not necessarily the "parent" molecule of cyanopolynnes. Instead, an enhanced cosmic ray flux, as found in the Galactic Center, would increase the abundance of ionised hydrocarbons, explaining the large HC$_3$N abundance and the cyanopolyynes ratios observed in G+0.693.

For the less stable isomers HCCNC and HNCCC, their abundances are systematically lower by a factor of 4 than those measured in TMC-1. HNCCC, the least stable isomer amongst the three molecules, presents the lowest abundance as expected. However, we found that the HCCCN : HCCNC : HNCCC ratios are almost uniform between G+0.693 and TMC-1 (see Table \ref{tab: abundance-ratio}). Recently, HCCNC and HNCCC have been detected in the prestellar core L1544 \citep{Vastel2018}. Together with previous detection of HC$_3$N by \citet{quenard_detection_2017}, the HCCCN : HCCNC : HNCCC abundance ratio toward L1544 is 1 : 0.04-0.14 : 0.003-0.01. This ratio is also marginally consistent with those obtained from G+0.693 and TMC-1.
\\
\\
\subsubsection{Nitrile, -CN group}
The closely related cyanides CH$_3$CN, C$_2$H$_3$CN, and C$_2$H$_5$CN have all been detected toward Sgr B2(N) and Orion KL, as well as G+0.693. Consistent molecular ratios are found between these three molecules toward Sgr B2N and Orion KL (i.e. 1 : 0.4 : 0.9 and 1 : 0.5 : 1.1 respectively), but a factor of 2 variation is present when compared to G+0.693 (1 : 0.8 : 0.4). Furthermore, a strong correlation has also been found for the abundances of C$_2$H$_3$CN and C$_2$H$_5$CN toward 6 hot molecular cores (HMCs), giving the abundance ratio C$_2$H$_3$CN : C$_2$H$_5$CN = 1 : $\sim$2-3.3 \citep{fontani_comparative_2007}. Similar ratios such as 1 : 2.25 and 1 : 2.2 are obtained toward Sgr B2N and Orion KL respectively. This is expected since hot cores are thought to have higher abundances of saturated molecules than those of unsaturated ones. Note here that the term "saturated" refers to carbon chain molecules involving one single C-C bond whereas the term "unsaturated" implies carbon chain molecules containing carbon-carbon double bonds (C=C) or triple bonds (C$\equiv$C). \citep{herbst_complex_2009}. And more importantly, C$_2$H$_5$CN in hot cores is thought to be formed onto icy mantles of dust grains via sequential hydrogenation of C$_3$N and then evaporated to form C$_2$H$_3$CN through ion-molecule reactions in the gas-phase \citep{Caselli1993}. On the contrary, unsaturated C$_2$H$_3$CN appears to be slightly more abundant than saturated C$_2$H$_5$CN in G+0.693, yielding an opposite ratio of C$_2$H$_3$CN : C$_2$H$_5$CN as 1 : 0.5 compared to hot cores. If we compare our results to the modelling of \citet{Caselli1993} for the Orion compact ridge, the ion-molecule gas phase chemistry can convert the saturated C$_2$H$_5$CN into unsaturated C$_2$H$_3$CN efficiently thanks to two different reasons: i) the high cosmic ray flux present in the GC that increases the fractional ionic abundance; and ii) the lower densities of $\sim$10$^4$ cm$^{-3}$ of G+0.693, which also are expected to yield higher ion densities in the gas \citep[see Section 3 in][]{Caselli1993}. In this way, for time-scales of about 10$^{5}$ years, C$_2$H$_3$CN can become more abundant than C$_2$H$_5$CN, which is consistent with our results. In addition, the CH$_3$CN/H$_2$ abundance ratio measured in G+0.693 matches very well with that of the shocked-region L1157-B1.
\\
\subsubsection{Amine, -NH/-NH$_2$ group}
\begin{figure*}
\includegraphics[width = \textwidth]{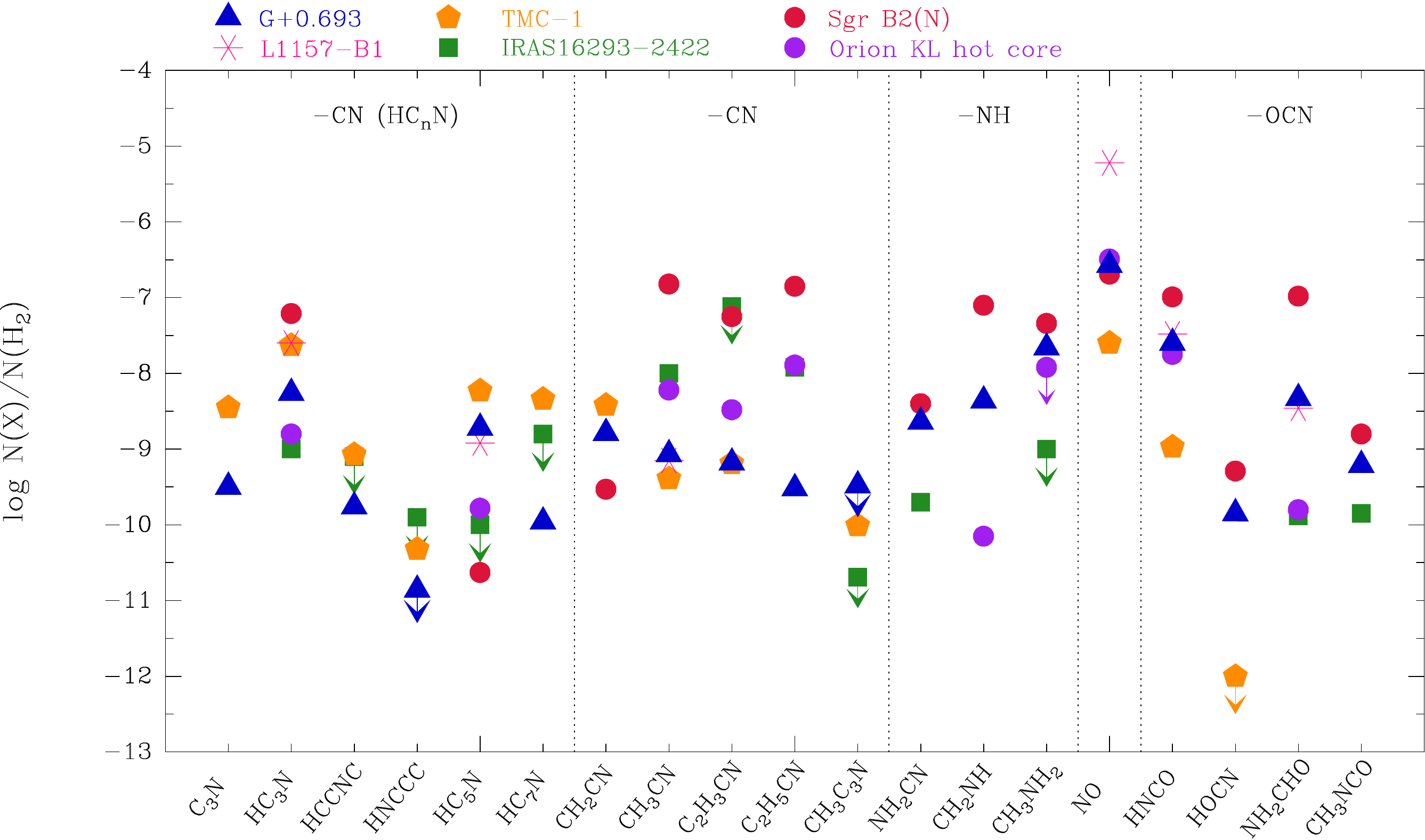}
\hspace{0.12cm}
 \caption{Molecular abundances relative to H$_2$ in different sources. Blue triangles represent the abundances derived from this work toward G+0.693; magenta stars are the results from L1157-B1 \citep{arce_complex_2008,mendoza_molecules_2014,Mendoza2018}; red circle are the results from Sgr B2N \citep{Ziurys1991,belloche_complex_2013,cernicharo_rigorous_2016}; purple circle represents the averaged results from Orion KL hot core \citep{blake_molecular_1987,sutton_distribution_1995}; orange pentagon and green rectangle represent values obtained from TMC-1 and IRAS16293 respectively \citep{Gerin1993,gratier_new_2016,jaber_census_2014,jaber_history_2017,Coutens2017,martin-domenech_detection_2017}.}
    \label{fig-fractional-abundance}
\end{figure*}

\begin{figure*}
   \begin{adjustbox}{addcode={\begin{minipage}{\width}}{\caption{
      Chemical diagram of N-bearing species to illustrated how each individual species can be related in different ways. Complexity increases from right to left by adding C ( four areas, C, C-C, C-C-C, and n-C divided by vertical line), from top to bottom by adding H, and addition of N is indicated in diagonal direction. Molecules are encapsulated in square boxes whilst radicals are encapsulated in ellipses. Dashed boxes indicate group of isomers of the same species. With relative abundance to H$_2$ provided beneath their molecular formula, species that have been detected in this study are in red boldface. Upper limits are also provided for undetected species. The rest of species that don't have relative abundance or upper limit provided are those have not been identified either due to lack of spectroscopic data or lack of sensitivity in observations are also indicated in G+0.693. 
      }{\label{fig-N-bearing-N}}\end{minipage}},rotate=90,center}
      \includegraphics[height=15cm, width=23cm]{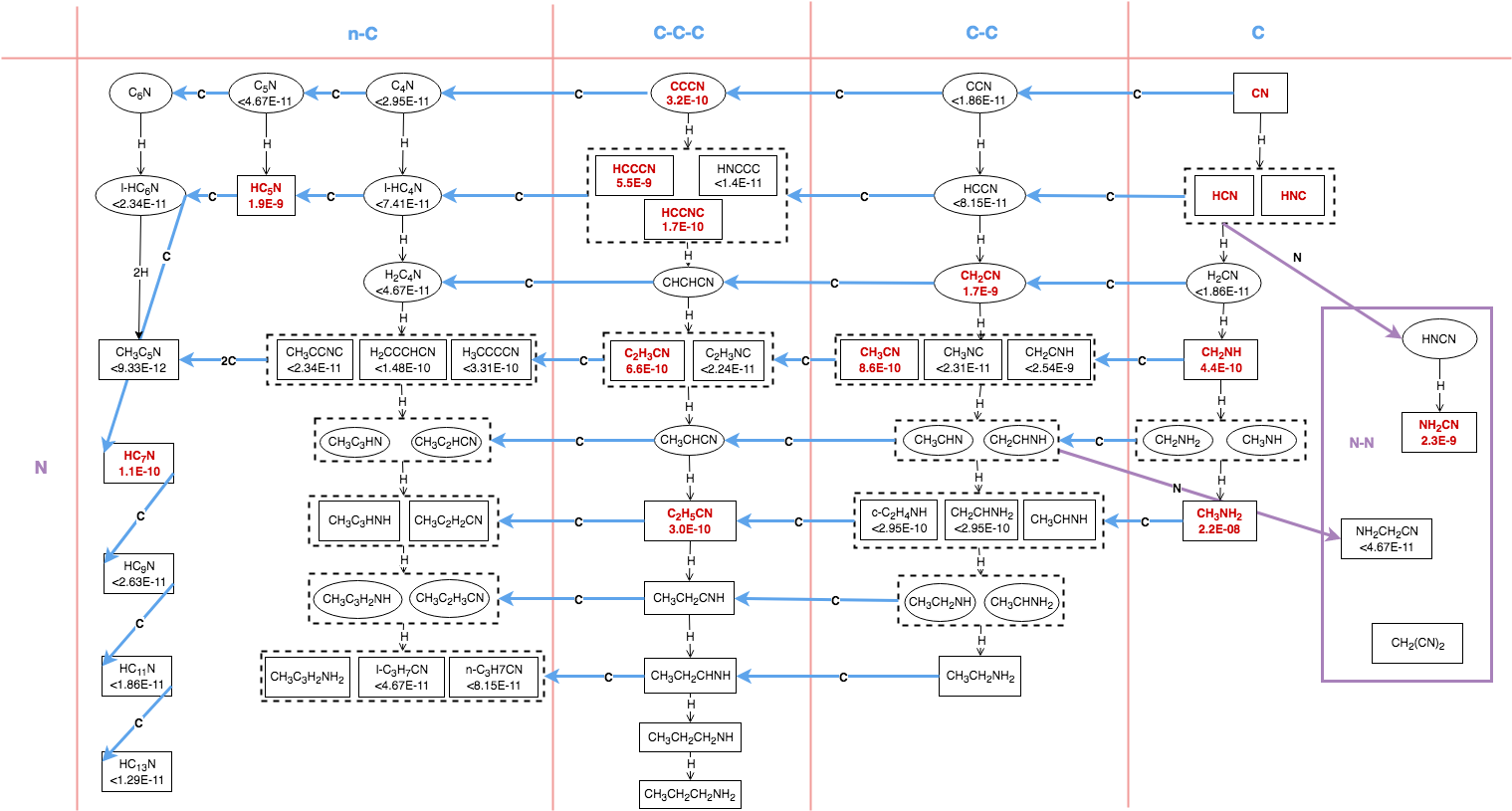}
  \end{adjustbox}
\end{figure*}

\begin{figure*}
\includegraphics[width = 0.9\textwidth]{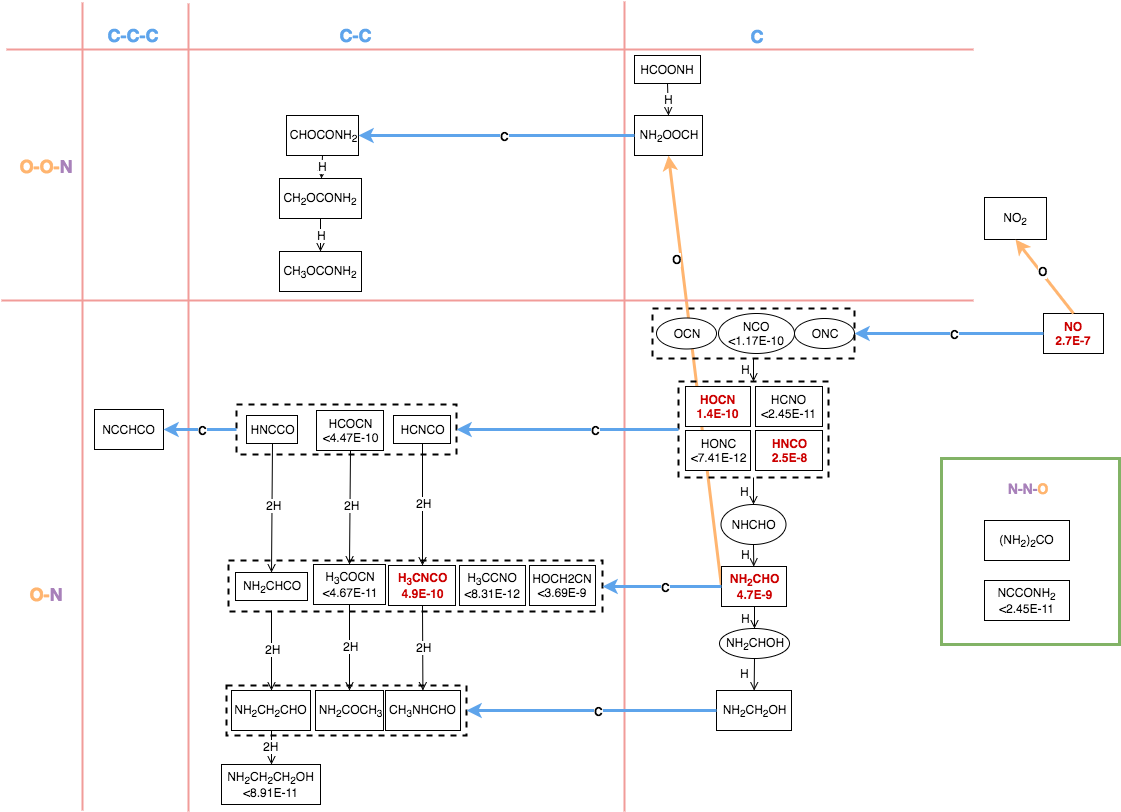}
%\hspace{0.12cm}
 \caption{Chemical diagram of O-containing N-bearing species that are potentially related. Same notations as used in Figure \ref{fig-N-bearing-N} except for O-addition instead of N-addition in diagonal direction.}
    \label{fig-N-bearing-O}
\end{figure*}

Among the studies we considered for our comparison, amino group species have mostly been detected in large abundance toward the high-mass star-forming region Sgr B2(N). Recently, NH$_2$CN has been reported by \citet{coutens_chemical_2017} toward IRAS16293 and NGC1333 IRAS2A hot corinos. Therefore, our knowledge of these amine group species has been restricted to regions with high temperatures and active star-formation. In this study, 4 different amine group species have been detected toward G+0.693, for which 3 out of 4 species present similar molecular abundances to those measured toward Sgr B2(N). The large abundance difference found for methanimine, (CH$_2$NH, see Figure \ref{fig-fractional-abundance}) can possibly be explained by the increased hydrogenation efficiency in G+0.693, which is expected to yield a higher abundance of methylamine, (CH$_3$NH$_2$). One of the possible formation routes of CH$_3$NH$_2$ is via hydrogenation of HCN on grain surfaces \citep{theule_hydrogenation_2011}. Grain surface hydrogenation of CH$_2$NH can also lead to the formation of CH$_3$NH$_2$ at low temperatures (T = 15$\,$K). Hence, CH$_2$NH is suggested to be a hydrogenation-intermediate species between HCN and CH$_3$NH$_2$ \citep{theule_hydrogenation_2011}. The abundance ratio of CH$_3$NH$_2$ : CH$_2$NH in G+0.693 is 1 : 0.1 whereas in the hot cores G10.47+00.3 and Sgr B2N, opposite ratios of 1 : 6 and 1 : 1.3 are obtained respectively \citep{ohishi_detection_2017,belloche_complex_2013}. Since CH$_3$NH$_2$ is likely formed via hydrogenation from HCN, the opposite trend for the CH$_3$NH$_2$ : CH$_2$NH ratio could be caused by a more efficient hydrogenation on dust grains in G+0.693 than in hot cores. In addition, coupled to the low dust temperatures measured in G+0.693, the chemistry in grain mantles is also likely to be affected by energetic processing in the GC such as X-rays and/or cosmic rays, which increases the availability of atomic H in the gas phase and subsequently, in the grain mantles. Indeed, \citet{armijos-abendano_3_2014} (source named as LOS+0.693 in the paper) has concluded that a chemistry driven by X-rays could be expected due to the presence of strong Fe K$\alpha$ line emission toward G+0.693, while UV photo-chemistry in G+0.693 is rather uncertain. Consequently, a larger degree of hydrogenation is possible in G+0.693 with respect to hot cores. Similarly, \citet{requena-torres_largest_2008} has shown that the abundance ratio of the O-bearing pair H$_2$CCO : CH$_3$CHO in G+0.693 (1 : 3.6) is completely different from that observed in hot cores (1 : 0.2). A more efficient hydrogenation in the CMZ with respect to hot cores was suggested in their study, which is consistent with our CH$_3$NH$_2$ : CH$_2$NH.

For the rest of amine species, similar abundances are derived in G+0.693 and Sgr B2(N). This suggests that these species may arise from the same environment, presumably the envelope of the Sgr B2 cloud. For other species within the same group, NH$_2$CN : NH$_2$CHO abundance ratio in IRAS16293 (1 : 5) is found to be an order of magnitude lower than in IRAS2A (1 : 50) where the latter falls within the range inferred toward Sgr B2(N) (1 : 25-50). For G+0.693, this ratio is found to be 1 : 2 which is consistent with the ratio measured in Orion KL (1 : $\sim$0.7-2.5) and about a factor of 2.5 lower than in IRAS16293. We speculate that the hydrogenation of NH$_2$ and CN radicals may be more efficient than the formation of NH$_2$CN since the enhanced cosmic ray flux is likely to produce larger amounts of atomic hydrogen, which are then available in the gas phase to accrete onto the surface of dust grains. 
\\
\\
\subsubsection{Cyanate, -NCO group}
Cyanate group molecules have attracted much attention in recent years due to their possible chemical link to the building blocks of life. N-C=O is often known as a peptide bond, which represents the linkage between two amino acids in protein chains. Several of these molecules with peptide-like bonds are detected in G+0.693 such as isocyanic acid (HNCO), formamide (NH$_2$CHO), and methyl isocyanate (CH$_3$NCO).

In the family of HNCO isomers, only the two most stable isomers HNCO and HOCN have been detected toward G+0.693 (HOCN was first observed in this source, named as SgrB2M offset 20'',100" in the paper by \citet{brunken_interstellar_2010}) and upper limits are provided for the other two isomers HCNO and HONC (see Table \ref{tab:non-detections}). HNCO in G+0.693 is almost as abundant as in L1157-B1 and only about a factor of 4 less abundant than in Sgr B2N. \citet{martin_tracing_2008} has proposed that shocks may explain the high HNCO abundances measured in galactic nuclei. Observations carried out toward the L1157 molecular outflow \citep{rodriguez-fernandez_hnco_2010}, also support the idea that the enhancement of HNCO in star-forming regions is due to shocks. Furthermore, two independent studies, \citet{Kelly_2016} and \citet{Yu2017}, have concluded that an enhancement of HNCO abundance may indicate the presence of a slow shock ($\sim$20$\,$km$^{-1}$).

On the whole, the consistent abundances of HNCO detected in these 3 regions imply that their physical environments are likely affected by low/moderate-velocity shocks. The fairly constant abundance ratio of HNCO : HOCN  ($\sim$1:0.005) between G+0.693 and Sgr B2N obtained in this study agrees very well with the HNCO : HOCN ratios derived from several chemically and physically distinct regions in Sgr B2 (see Table 3. in \citet{brunken_interstellar_2010}). This further supports the hypothesis that, like HNCO, sputtering of grain mantles by shock waves are responsible for the observed abundance of HOCN in these regions. 

The HNCO : HOCN : HCNO ratio behaves differently in G+0.693, gives the ratio 1 : 0.006 : $\leq$0.001 compared to that in the proto-star IRAS16293 (1 : $\leq$0.0002 : $\leq$0.0007) and in the prestellar core L1544 \citep[1 : 0.03 : 0.02, methanol peak; see][]{quenard_chemical_2017}. This might be potentially related to the destruction rates of HOCN and HCNO via the gas-phase reactions HOCN/HCNO + O, since both are sensitive to the gas temperature \citep{quenard_chemical_2017}. Their reaction rates increase rapidly from 10$^{-18}$ to 10$^{-10}$ cm$^3$ s$^{-1}$ for temperatures from 10 to 300$\,$K, which would yield more extreme HNCO : HOCN : HCNO ratios in IRAS16293. In this context, with a gas temperature of $\sim$100-150$\,$K, the destruction rates of HOCN and HCNO in G+0.693 are not as efficient as those in IRAS16293 but certainly faster than in L1544. Consequently the HNCO : HOCN : HCNO ratio in G+0.693 is expected to lie within the corresponding ratios obtained from these two regions, as observed.

Besides the family of HNCO isomers, NH$_2$CHO and HNCO have been proposed to be chemically related \citep{mendoza_molecules_2014,lopez-sepulcre_shedding_2015}. 
However, recent modelling by \citet{quenard_chemical_2017} has shown that these two species, rather than being chemically linked, respond in the same manner to environmental conditions, precisely to the temperature. This is consistent with the measured HNCO : NH$_2$CHO abundance ratio in G+0.693 (1 : 0.2), which lies between those derived toward the prestellar core L1544 (methanol peak, 1 : $\leq$0.03) and the hot corino IRAS16293 (1 : 1). Note that the gas temperature progressively increases from L1544 ($\sim$10$\,$K) to G+0.693 ($\leq$145$\,$K), and to the IRAS16293 hot corino ($\sim$100$\,$ to up to 300$\,$K). The measured abundance of NH$_2$CHO in G+0.693 is much higher than those observed in both IRAS16293 \citep[$\leq$1.9$\times$ 10$^{-9}$,][]{martin-domenech_detection_2017} and L1544 \citep[$\leq$(6.7-8.7)$\times$ 10$^{-13}$,][]{jimenez-serra_spatial_2016}, but very similar to that inferred from the L1157 molecular outflow \citep{mendoza_molecules_2014}. It is possible that NH$_2$CHO is formed either in shocks via gas-phase reactions \citep{codella_seeds_2017} or that it is formed on grain surfaces and then ejected from dust grains in shocks \citep{quenard_chemical_2017}.

In Table \ref{tab: abundance-ratio}, we present the abundance ratio between HNCO, NH$_2$CHO and other related molecules such as NCO and CH$_3$NCO. The latter molecule, considered as the relevant precursor in the formation of prebiotic species, have recently been detected in several sources such as Sgr B2N \citep{halfen_interstellar_2015,belloche_rotational_2017}, Orion KL \citep{cernicharo_rigorous_2016}, IRAS16293 \citep{ligterink_alma-pils_2017,martin-domenech_detection_2017} and even in comet 67P/Churyumov-Gerasimenko \citep{Goesmann2015}. We find that the HNCO : CH$_3$NCO ratio in G+0.693 is a factor of 4, 5.5, and 3.5 lower than those measured in IRAS16293, Sgr B2N and Orion KL respectively. In contrast, the NH$_2$CHO : CH$_3$NCO ratio is a factor of 1.3 and 3.3 higher than those measured in IRAS16293 and Sgr B2N respectively but a factor of 35 lower than in Orion KL. The measured discrepancies in the abundance ratio between these regions point to a different chemical evolution of the grain mantles in the GC and in the disk see also \citep[see also][]{requena-torres_largest_2008}.

%%%%%%%%%%%%%%%%%%%%%%%%%%%%%%%%%%%%%%%%%%
%%%%%%% FIGURE-FRACTIONAL-ABUNDANCE %%%%%%
%%%%%%%%%%%%%%%%%%%%%%%%%%%%%%%%%%%%%%%%%%
%%%%%%%%%%%%%%%%%%%%%%%%%%%%%%%%%%%%%%%%%%%%%%%%%%

%%%%%%%%%%%%%%%%%%%%%%%%%%%%%%%%%%%%%%%%%%
%%%%%%%% TABLE-ABUNDANCE-RELATED %%%%%%%%%
%%%%%%%%%%%%%%%%%%%%%%%%%%%%%%%%%%%%%%%%%%

\begin{table*}
  \centering
  \caption{Observational molecular abundances between related molecules}
   %\begin{tabular}{|p{3.5cm}|p{3.5cm}|}
   \begin{adjustbox}{width=\textwidth}
   \begin{tabular}{llllll}
   \hline
   \hline
   Molecules & \multicolumn{5}{c}{Abundance ratios} \\
   \hline
   \multicolumn{6}{c}{Isomers} \\
   \hline
    & G+0.693 & SgrB2(N) & Orion KL & IRAS16293 & TMC-1\\
   \hline
   %CH$_3$CN : CH$_2$CNH : CH$_3$NC & 1 : $\leq$3 : 0.03 & 1: 0.02 & - & 1 : $\leq$0.1 & - \\
   HCCCN : HCCNC : HNCCC & 1 : 0.03 : 0.003 & - & -  & 1 : $\leq$0.8 : $\leq$0.1 & 1 : 0.04 : 0.002\\
   %CH$_3$C$_3$N : H$_2$CCCHCN : CH$_3$CCNC & 1 : $\leq$0.4 : $\leq$0.07 & -  & - & - & - \\
   HNCO : HOCN : HCNO : HONC & 1 : 0.006 : $\leq$0.001 : $\leq$0.0003 & 1 : 0.005 & - & 1 : $\leq$0.0002 : $\leq$0.0007$^{a}$ & 1 : $\leq$0.0008 \\
   CH$_3$NCO : CH$_3$OCN : CH$_3$CNO & 1 : $\leq$0.1 : $\leq$0.02 & 1  : - : $\leq$0.01  & - & 1 : $\leq$0.1 : $\leq$0.01$^{a}$ & - \\
   \hline
   \multicolumn{6}{c}{Parents $\&$ Daughters Species} \\
   \hline
   HC$_3$N : HC$_5$N : HC$_7$N : HC$_9$N & 1 : 0.3 : 0.02 : $\leq$0.005 & 1 : 0.004 : - : - & 1 : 0.1 : - : - & 1 : 0.09 : $\leq$1 & 1 : 0.25 : 0.19 : 0.04 \\
   CH$_3$CN : C$_2$H$_3$CN : C$_2$H$_5$CN : n-C$_3$H$_7$CN & 1 : 0.8 : 0.4 : $\leq$0.01 & 1 : 0.4 : 0.9 : 0.008 & 1 : 0.5 : 1.1 & 1 : $\leq$8 : 1.2 : $\leq$0.05 & 1: 1.6 \\
   CH$_3$CN : CH$_3$C$_3$N : CH$_3$C$_5$N & 1 : 0.4 : $\leq$0.01 & - & - & 1 : $\leq$0.002 : $\leq$0.01 & 1: 0.2 \\
   CH$_3$NH$_2$ : CH$_2$NH & 1 : 0.1 & 1 : 1.3 & - & - & - \\
   HNCO : NH$_2$CHO : CH$_3$NCO & 1 : 0.2 : 0.02 & 1 : 1 : 0.03 & 1 : 0.02 : 0.07 & 1 : 1 : 0.08 & - \\ 

   \hline
   \hline
  \end{tabular}
  \end{adjustbox}
  \begin{flushleft}
  (a) Upper limit estimated from chemical model in \citet{quenard_chemical_2017}\\
  \end{flushleft}
  \label{tab: abundance-ratio}
\end{table*}

%%%%%%%%%%%%%%%%%%%%%%%%%%%%%%%%%%%%%%%%%%
%%%%%%%% FIGURE-N-BEARING %%%%%%%%%%%%%%%%
%%%%%%%%%%%%%%%%%%%%%%%%%%%%%%%%%%%%%%%%%%

%%%%%%%%%%%%%%%%%%%%%%%%%%%%%%%%%%%%%%%%

\subsection{Comparison with extragalactic environments}
The QGMC G+0.693 in the GC is not only one of the most promising sources to constrain the formation pathway of COMs detected in the interstellar medium, but also an excellent candidate to enhance our knowledge of the chemical complexity expected in the nuclei of external galaxies and the influence of the different extragalactic sources. There is an increasing number of N-bearing species being detected across different extragalactic environments, evidencing similarities as well as differences compared to galactic environments. For example, the first detection of NO in the starburst galaxy NGC 253 by \citep{martin2003,Martin2006} presents an abundance consistent to that measured in G+0.693, Sgr B2N and Orion KL. The rare molecule NH$_2$CN has also been detected toward NGC 253 with an abundance of $\sim$2 $\times$10$^{-10}$ \citep[][]{Martin2006,Aladro2015}. This abundance is an order of magnitude lower than that in Sgr B2N and G+0.693 but matches up with the latest measurement toward IRAS16293 \citep{Coutens2017}. HCN, HNC, CH$_3$CN, HNCO, and HC$_3$N have also been detected in several extragalactic sources \citep{Martin2006,Aladro2011,Aladro2015,Costaliola2015,Harada2018}. 
 
The most accepted idea is that the chemistry of the most complex molecules in starburst galaxies is dominated by the ejection of the icy grain mantel by low velocity shock followed by different types of gas phase chemistry depending on the dominant environment (UV radiation, X-rays, cosmic rays, high gas temperature etc.). One of the most interesting results from our study of G+0.693 is that one can use the HC$_3$N/HC$_5$N abundance ratio as a tracer of enhanced cosmic ray fluxes. So far, larger cyanopolyynes such as HC$_5$N, however, have only been detected in two extragalactic sources: NGC 253 \citep{Aladro2015} and the luminous infrared galaxy NGC 4418 \citep{Costaliola2015}. Their derived HC$_3$N : HC$_5$N column density ratios are 1 : 0.7 and 1 : 0.6 respectively, i.e. at least factors of 3 higher than those measured in any other galactic source (see Table 7). These large ratios would be consistent with the scenario of enhanced cosmic rays. This would indicate that the nuclei of both active galaxies would have a cosmic ray flux even larger than that inferred for the GC. For both source similar conclusion has been reached using other tracers. Using the emission from OH$^+$, H$_2$O$^+$, and H$_3$O$^+$, \citep{Gonzalex-Alfonso2012} concluded that X-rays/cosmic ray ionization from the AGN is very likely responsible for the large abundance of these molecules. \citep{Bradford2003} have analysed the emission of high J-line of CO and concluded that the best mechanism for heating the gas is cosmic rays with a flux 800 times that of the Galaxy. 

\subsection{The origin of N-bearing species in G+0.693}

The QGMC G+0.693 is located toward the north-east of the Sgr B2 star-forming complex but it does not show any signposts of recent or ongoing star formation in the form of UC HII regions or H$_2$O masers \citep{gusten_ammonia_1981,huettemeister_kinetic_1993,martin-pintado_sio_1997,ginsburg2018}. Same as other QGMCs within the CMZ, its physical properties are characterised by high gas kinetic temperatures and cold dust temperatures \citep{Guesten1985,Rodriguez-Fernandez2001,Rodriguez-Fernandez2004}. As discussed throughout Section 4.1, the chemistry of this QGMC is rather unique since it shows high abundances not only of O-bearing species \citep{requena-torres_largest_2008} but also of N-bearing molecules. The reasons why this source presents such a rich chemistry in COMs among all QGMCs in the Galactic Center, remains however unclear. 

The comparison of the abundances of N-bearing COMs measured in G+0.693 with those derived toward other Galactic environments (Section 4.1), provides clues about the main mechanism(s) responsible for the rich COM chemistry found in this source:

\begin{itemize}
  \item Large carbon chain cyanopolyynes (from HC$_3$N to HC$_7$N) are clearly detected toward G+0.693. Systematically lower abundances of HC$_3$N, HC$_5$N, and HC$_7$N are found in G+0.693 compared to the molecular dark cloud TMC-1. This indicates that CN might not necessarily be the "parent" molecule of cyanopolyynes. In addition, the lower HC$_5$N : HC$_7$N ratios observed in G+0.693 may be due to the presence of enhanced cosmic-rays ionisation rates in the Galactic Center.
  \item In G+0.693, the saturated molecule CH$_3$NH$_2$ appears to be more abundant than the unsaturated radical CH$_2$NH whilst the opposite is true in Galactic hot cores. A more efficient hydrogenation mechanism on dust grains is proposed for G+0.693 than for hot cores in the Galactic disk. The most likely scenario to explain the large abundance of unsaturated C$_2$H$_3$CN compared with that of the saturated C$_2$H$_5$CN is ion-molecule gas phase chemistry fostered by an enhanced cosmic ray flux.
  \item Comparison of the abundance ratio of peptide-like (-CNO group) species across multiple environments shows large differences, which supports the previous claim in \citet{requena-torres_largest_2008} of a different grain mantle composition due to a different chemical evolution of the grain mantles in the GC with respect to the Galactic disk.
  \item Remarkable consistences between the molecular abundance measured in G+0.693 and in the shocked region L1157-B1 (e.g. CH$_3$CN, HC$_5$N, HNCO and NH$_2$CHO) are found. This emphasises the idea that a large fraction of the ices from dust grains has been injected into the gas phase via grain sputtering in widespread low-velocity shock waves. 
\end{itemize} 

It has been proposed that G+0.693 is located between two streams of molecular gas that seem to be merging \citep{hasegawa_large-scale_1994,henshaw_molecular_2016}. This may yield a cloud-cloud collision that drives large-scale, low-velocity shocks in the region, and which ultimately sputters dust grains icy mantles efficiently. This QGMC indeed shows the highest abundances of HNCO in the sample of \citet[][source SgrB2M offset 20",100" in that paper]{martin_tracing_2008}, supporting the idea that this source is mainly dominated by low-velocity shock. This would also explain the similarity in the abundances of N-bearing species measured toward G+0.693 and L1157-B1 (see Section 4.1). Some of the derived abundance ratios (such as e.g. HC$_5$N : HC$_7$N, and the high abundance of C$_2$H$_3$CN) indicate that energetic processing by e.g. X-rays and/or cosmic rays may strongly affect the gas phase chemistry of this cloud owing to its location in the CMZ. 

%%%%%%%%%%%%%%%%%%%% CONCLUSIONS %%%%%%%%%%%%%%%%%%

\section{Conclusions}
Using the GBT and IRAM 30$\,$m telescopes, we performed an unbiased spectral line survey toward the GC QGMC G+0.693. The survey covers partially the 1cm and 1mm spectral window, and fully the 2mm, and 3mm atmospheric windows. We explore the chemical richness in terms of presence and abundance of N-bearing species in the GC QGMC G+0.693. In this study, we have reported 17 clear detections and 2 tentative detections of N-bearing species. These species show very high abundances relative to H$_2$, ranging from 10$^{-11}$ to 10$^{-8}$. The comparison across various galactic environments allow us to constrain possible mechanisms responsible for the unique chemistry observed in G+0.693: grain sputtering by widespread low-velocity shocks is by far the most promising mechanism to activate the chemistry in this source. However, energetic processing by either X-rays and/or cosmic rays needs to be invoked in order to explain some of the observed abundance ratios between molecules within the same family. Partial contribution from gas-phase chemistry cannot be ruled out either for some particular cases. The comparison of the measured molecular abundance in G+0.693 with those derived in extragalactic sources, shows that G+0.693 is an excellent template where to elucidate the chemical complexity expected in future extragalactic surveys done with ALMA. 

Although the current data have highlighted an extremely rich organic inventory in G+0.69 with abundant amounts of complex N-bearing species, making it to be one of the the largest organic species repository in the CMZ, the nature of this source has not yet been unveiled in detail. Interferometric maps are urged to establish the morphology and small-scale physical structure and thus understand fundamentally the chemical stratification in this source. New observations with greater sensitivity and higher angular resolution, as well as more laboratory experiments and theoretical models, are crucial to investigate further the origin of COMs not only in G+0.693 but also in extragalactic sources.

%%%%%%%%%%%%%%%%%%%%%%%%%%%%%%%%%%%%%%%%%%%%

%%%%%%%%%%%%%%%% ACKNOWLEDGEMENTS %%%%%%%%%%%%%%%%

\section*{Acknowledgements}
We thank the anonymous referee for his/her instructive comments
and suggestions. S. Z acknowledges support through a Principal's studentship funded by Queen Mary University of London. I.J.-S. acknowledges the financial support received from the STFC through an Ernest Rutherford Fellowship (proposal number ST/L004801). V.M.R. has received funding from the European Union's H2020 research and innovation programme under the Marie Sk\l{}odowska-Curie grant agreement No 664931. J.M.-P. has been partially supported by the Spanish MINECO under grant numbers: ESP2015-65597-C4-1-R. and ESP2017-86582-C4-1-R. D.R. acknowledges support from the Collaborative Research Council 956, subproject A5, funded by the Deutsche Forschungsgemeinschaft (DFG).

%%%%%%%%%%%%%%%%%%%%%%%%%%%%%%%%%%%%%%%%%%%%%%%%%%


%\cleardoublepage
\bibliographystyle{mnras}
\bibliography{bibliography} % if your bibtex file is called example.bib

%\begin{thebibliography}{99}
%\end{thebibliography}

%%%%%%%%%%%%%%%%%%%%%%%%%%%%%%%%%%%%%%%%%%%%%%%%%%

%%%%%%%%%%%%%%%%% APPENDICES %%%%%%%%%%%%%%%%%%%%%
%%%%%%%%%%% MOLECULAR SPECTROSCOPY %%%%%%%%%%%%%%%
\appendix
\section{Molecular spectroscopy}
\subsection{Methyl cyanide CH\texorpdfstring{$_3$},CN}
CH$_3$CN is a strongly prolate symmetric tops molecule, whose rotational energy levels can be described by total angular momentum, J, and its projection along the symmetry axis, K. The permitted radiative transitions according to the selection rules are only within K ladders i.e. $\Delta$K = 0. Therefore, successive K-components of a particular J $\rightarrow$ (J - 1) transition are expected to be present simultaneously in a narrow frequency band. In addition, the presence of the identical hydrogen nuclei gives rise to two symmetry species, distinguished by their nuclear spin state, denoted A- and E-state. Energy levels with K = 3n belong to A-states while E-states are referred to K $\neq$ 3n with n$\geq$0 \citep{boucher_high-resolution_1977,cazzoli_lamb-dip_2006,muller_rotational_2009}. In general, the statistical weight between A- and E-states are 2:1. However, if both states are formed in equilibrium conditions, the abundance ratios is expected to be ~1 \citep{minh_measurement_1993}.

%%%%%%%%%%%%%%%%%%%%%%%%%%%%%%%%%%%%%%%%%%

\subsection{Cyanomethyl radical H\texorpdfstring{$_2$},CCN and cyamamide NH\texorpdfstring{$_2$},CN}
The radical CH$_2$CN is the simplest cyanide derivative of the methyl radical, CH$_3$. It has two interchangeable hydrogen nuclei with non-zero spin which dictate the existence of an ortho-para symmetry. The same occurs in the NCN-frame contained molecule, NH$_2$CN. For both species, the quantum number K$_a$ is even for ortho levels and odd for para levels. The statistical weight of ortho : para is 3 : 1 due to their nuclear spin degeneracies, this means transitions of ortho levels would be three times stronger than equivalent transition of the para form \citep{millen_microwave_1962,flower_ortho-h2/para-h2_1984,takakuwa_ortho--para_2001}. 

%%%%%%%%%%%%%%%%%%%%%%%%%%%%%%%%%%%%%%%%%%

\subsection{Methanimine CH\texorpdfstring{$_2$},NH, vinl cyanide C\texorpdfstring{$_2$},H\texorpdfstring{$_3$},CN, ethyl cyanide C\texorpdfstring{$_2$},H\texorpdfstring{$_5$},CN, isocyanic acid HNCO, formamide NH\texorpdfstring{$_2$}CHO, and methylamine CH\texorpdfstring{$_3$},NH\texorpdfstring{$_2$},}
CH$_2$NH is a near prolate planar asymmetric rotor, where the components of the electric dipole moment are constrained to lie along the \textit{a} and \textit{b} principal axes. This results in two types of allowed transitions in the rotational spectrum: a-type, in which the K$_a$ (prolate) quantum number does not change and the K$_b$ (oblate) quantum number changes by one unit (i.e. $\Delta$K$_a$=0, $\Delta$K$_b$=$\pm$1), and b-type, in which both K$_a$ and K$_c$ change by one unit (i.e. $\Delta$K$_a$=$\pm$1, $\Delta$K$_b$=$\pm$1), \citep{kirchhoff_microwave_1973,dore_magnetic_2010,dore_accurate_2012}. 

C$_2$H$_3$CN and C$_2$H$_5$CN are planar asymmetric rotors with a- and b-type transitions allowed in the rotational spectrum \citep{neill_herschel_2014}. All transitions detected in G+0.693 are stronger a-type spectra. 

HNCO and NH$_2$CHO are simple asymmetric prolate rotors, hence a- and b-type transitions are allowed. For HNCO, only a-type transitions within the K$_a$=0 and 1 ladders are detected. For NH$_2$CHO, a-type transitions are observed from each K$_a$ ladder up to K$_a$=3 whilst only K$_a$=0 and 1 ladders are observed in b-type transitions.

CH$_3$NH$_2$ is a near-prolate asymmetric top molecule, whose a- and b-type transitions are allowed. In this case, both a-type and b-type transitions within the K$_a$=0 and 1 ladders are detected.
%%%%%%%%%%%%%%%%%%%%%%%%%%%%%%%%%%%%%%%%%%

%%%%%%%%%%% LINE PROFILES %%%%%%%%%%%%%%%
\section{Spectra of detected molecules}
\begin{figure*}
\includegraphics[width = 0.7\textwidth]{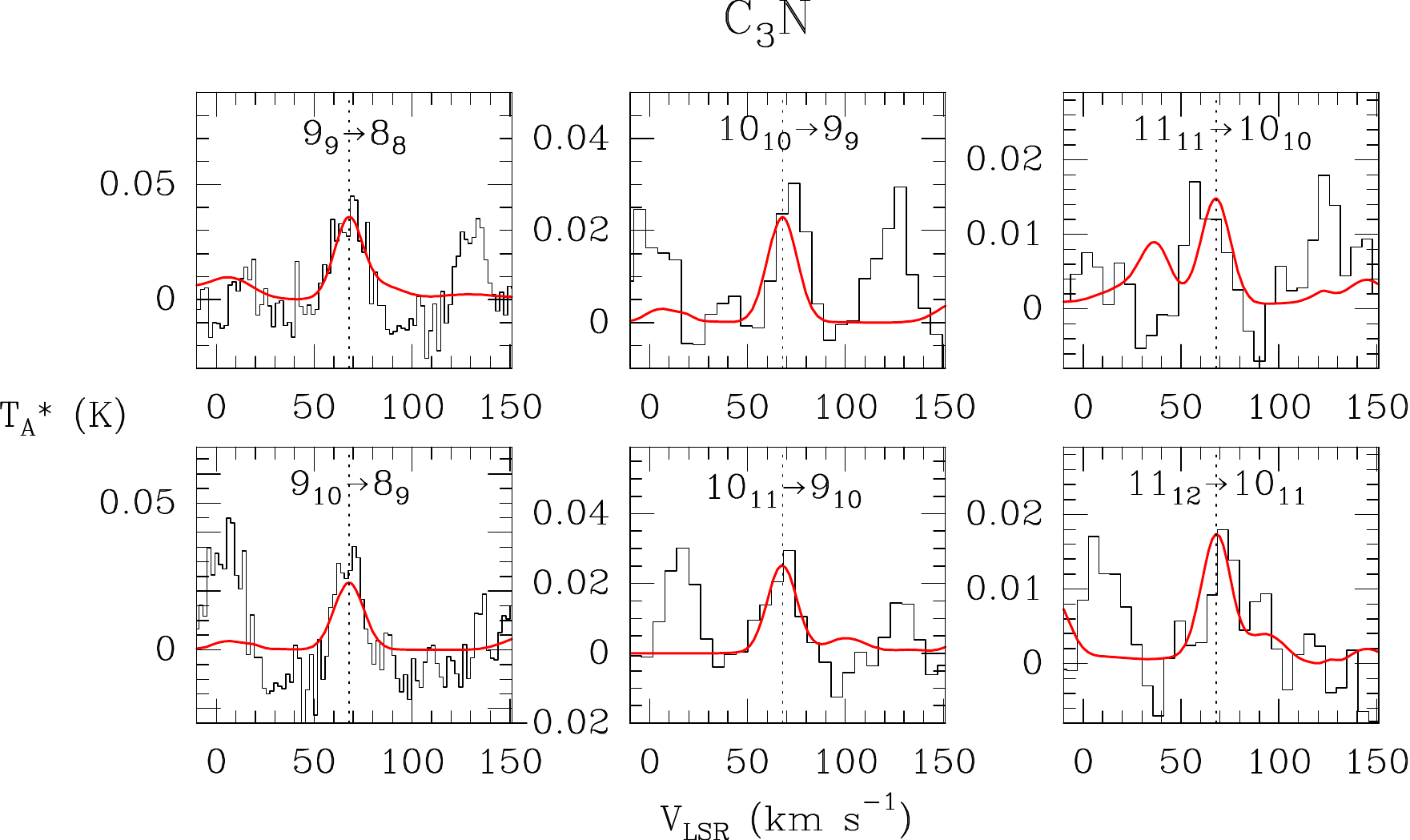}
 \caption{C$_3$N detected lines (black histogram spectra). The LTE best fits from MADCUBA are also shown overlaid in red. The vertical dotted lines correspond to the central velocity of the G+0.693 cloud. Note that other fitted lines that are not aligned with the dotted lines are arose from other species.}
    \label{fig-C3N}
\end{figure*}
%%%%%%%%%%%%%%%%%%%%%%%%%%%%%%%%%%%%%%%%%%
\begin{figure*}
\includegraphics[width = 0.7\textwidth]{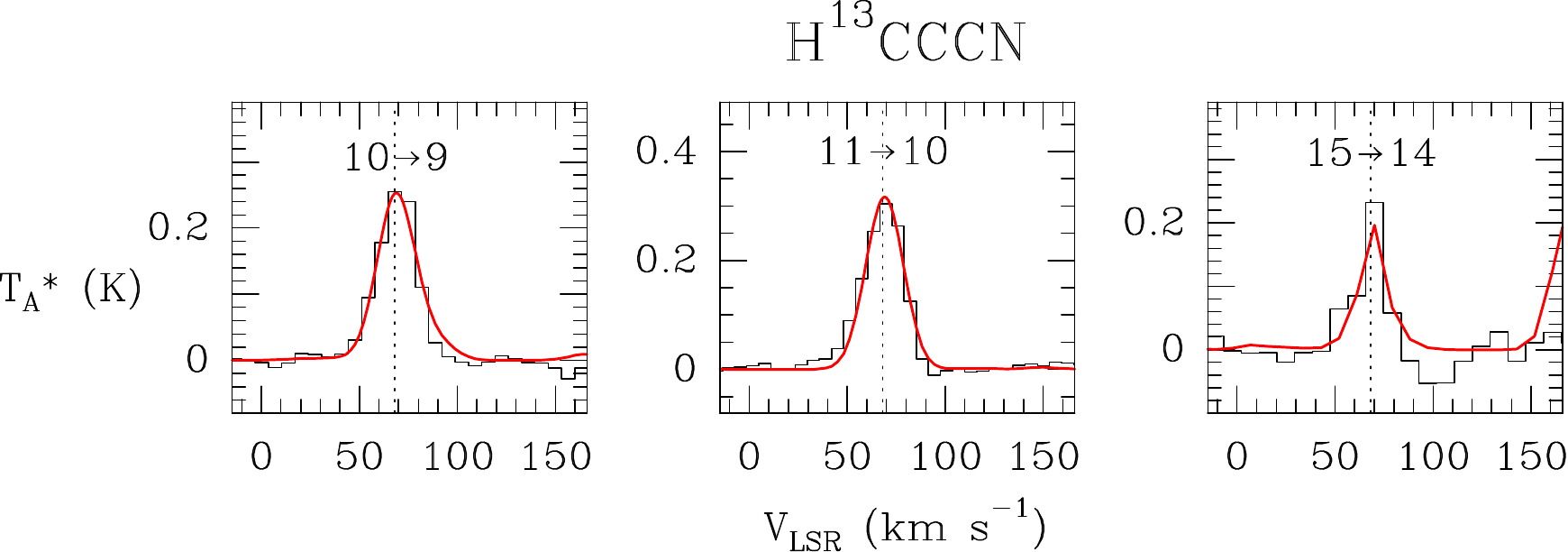}
 \caption{H$^{13}$CCCN detected lines (black histogram spectra). The LTE best fits from MADCUBA are also shown overlaid in red. The vertical dotted lines correspond to the central velocity of the G+0.693 cloud. Note that other fitted lines that are not aligned with the dotted lines are arose from other species.}
    \label{fig-HC3N}
\end{figure*}
%%%%%%%%%%%%%%%%%%%%%%%%%%%%%%%%%%%%%%%%%%
\begin{figure*}
\includegraphics[width = 0.7\textwidth]{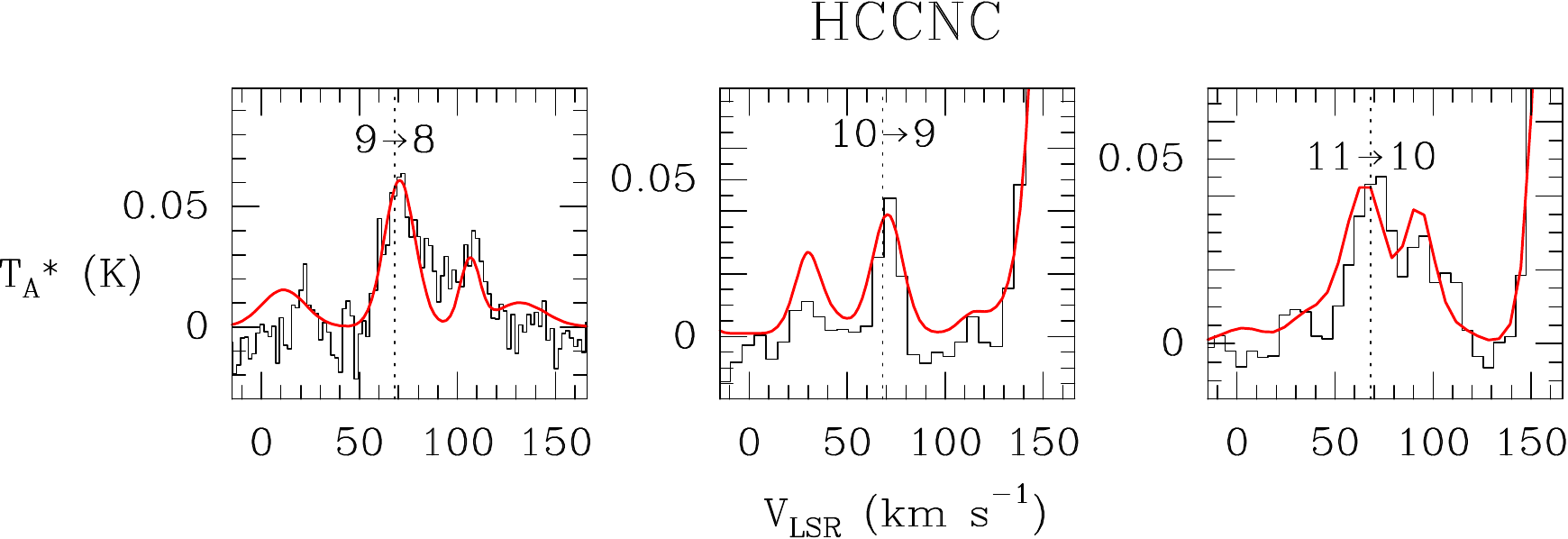}
 \caption{HCCNC detected lines (black histogram spectra). The LTE best fits from MADCUBA are also shown overlaid in red. The vertical dotted lines correspond to the central velocity of the G+0.693 cloud. Note that other fitted lines that are not aligned with the dotted lines are arose from other species.}
    \label{fig-HCCNC}
\end{figure*}
%%%%%%%%%%%%%%%%%%%%%%%%%%%%%%%%%%%%%%%%%%
\begin{figure*}
\includegraphics[width = 0.85\textwidth]{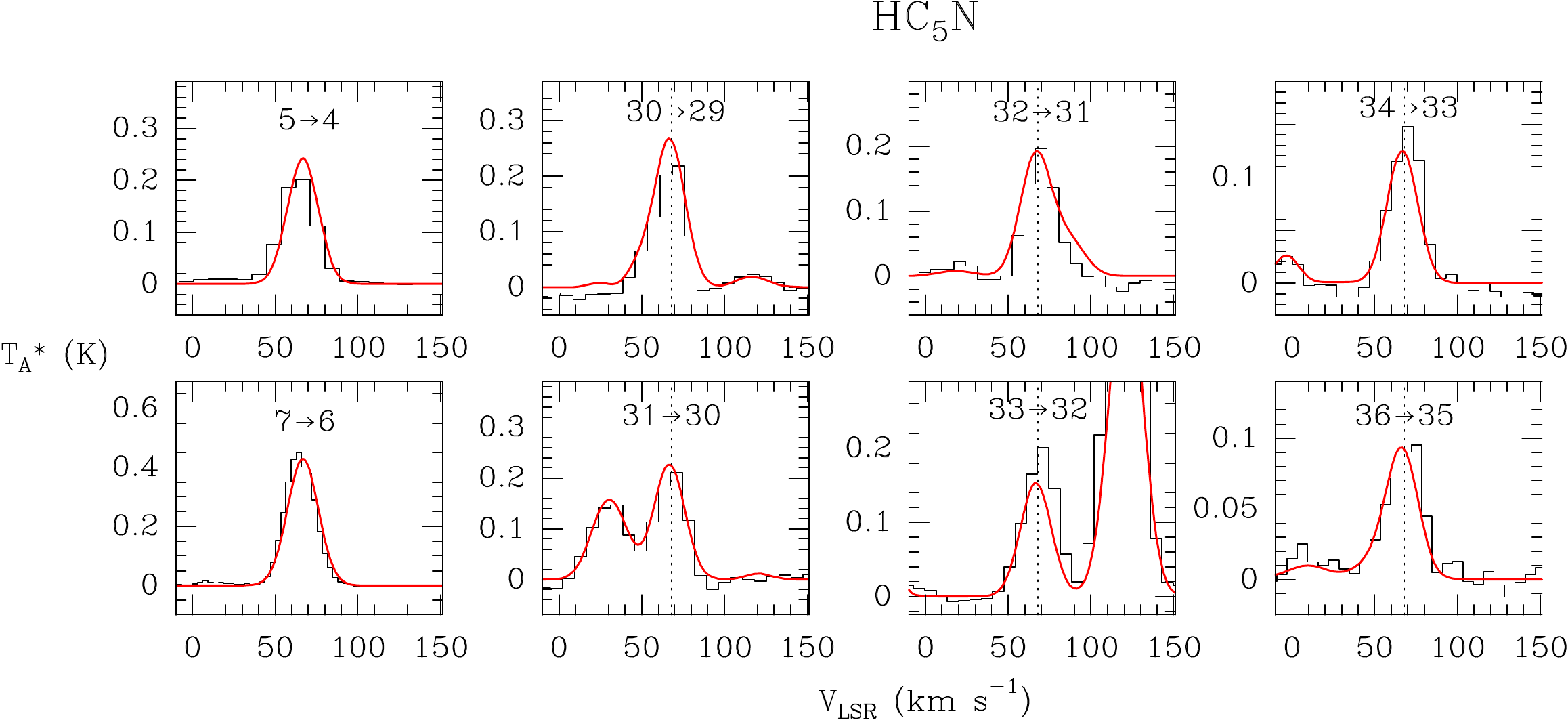}
 \caption{HC$_5$N detected lines (black histogram spectra). The LTE best fits from MADCUBA are also shown overlaid in red. The vertical dotted lines correspond to the central velocity of the G+0.693 cloud. Note that other fitted lines that are not aligned with the dotted lines are arose from other species.}
    \label{fig-HC5N}
\end{figure*}
%%%%%%%%%%%%%%%%%%%%%%%%%%%%%%%%%%%%%%%%%%
\begin{figure*}
\includegraphics[width = \textwidth]{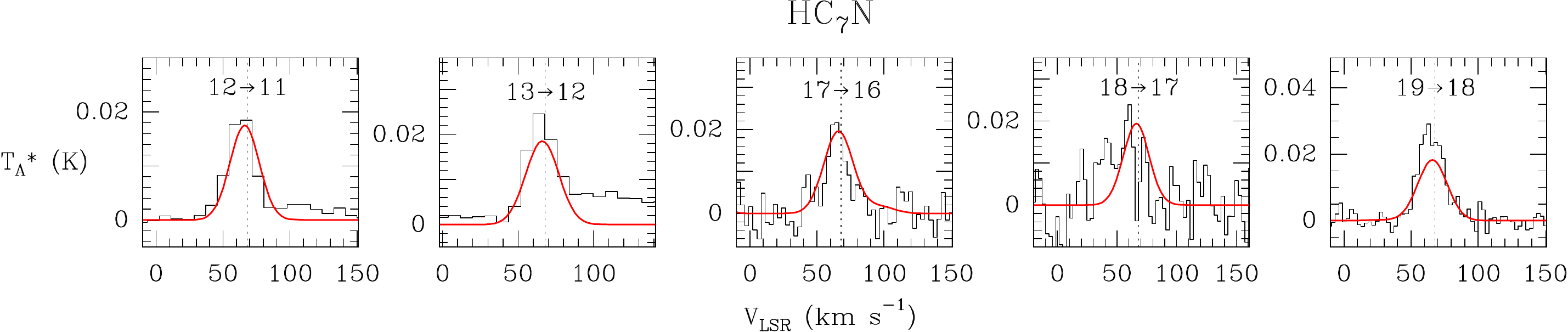}
 \caption{HC$_7$N detected lines (black histogram spectra). The LTE best fits from MADCUBA are also shown overlaid in red. The vertical dotted lines correspond to the central velocity of the G+0.693 cloud. Note that other fitted lines that are not aligned with the dotted lines are arose from other species.}
    \label{fig-HC7N}
\end{figure*}
%%%%%%%%%%%%%%%%%%%%%%%%%%%%%%%%%%%%%%%%%%
\begin{figure*}
\includegraphics[width = 0.85\textwidth]{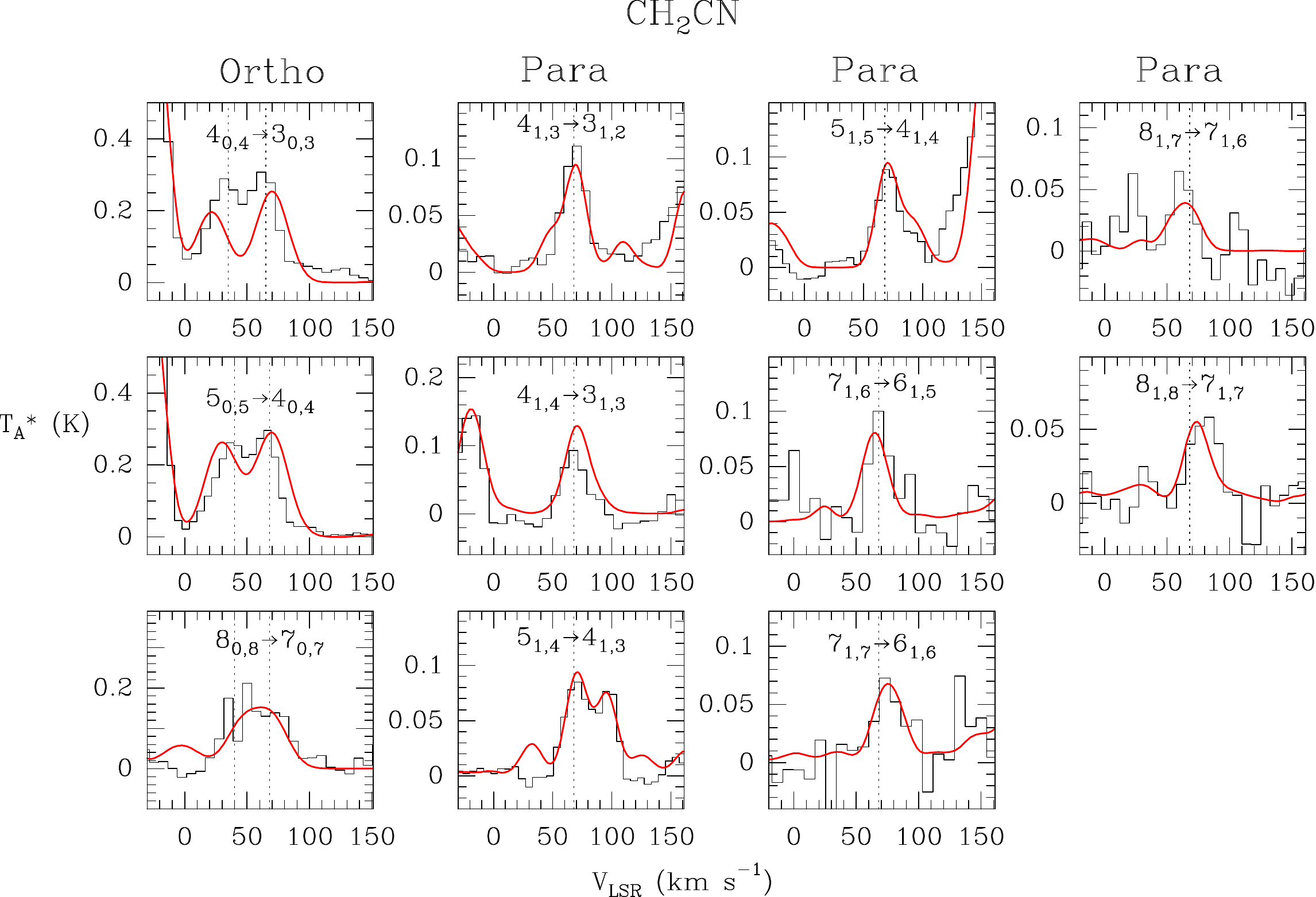}
 \caption{CH$_2$CN detected lines (black histogram spectra). The LTE best fits from MADCUBA are also shown overlaid in red. For ortho-CH$_2$CN, the two vertical dotted lines indicate the nearby hyperfine structure components. Whilst for para-Ch$_2$CN, the vertical dotted lines correspond to the central velocity of the G+0.693 cloud. Note that other fitted lines that are not aligned with the dotted lines are arose from other species.}
    \label{fig-CH2CN}
\end{figure*}
%%%%%%%%%%%%%%%%%%%%%%%%%%%%%%%%%%%%%%%%%%
\begin{figure*}
\includegraphics[width = 0.85\textwidth]{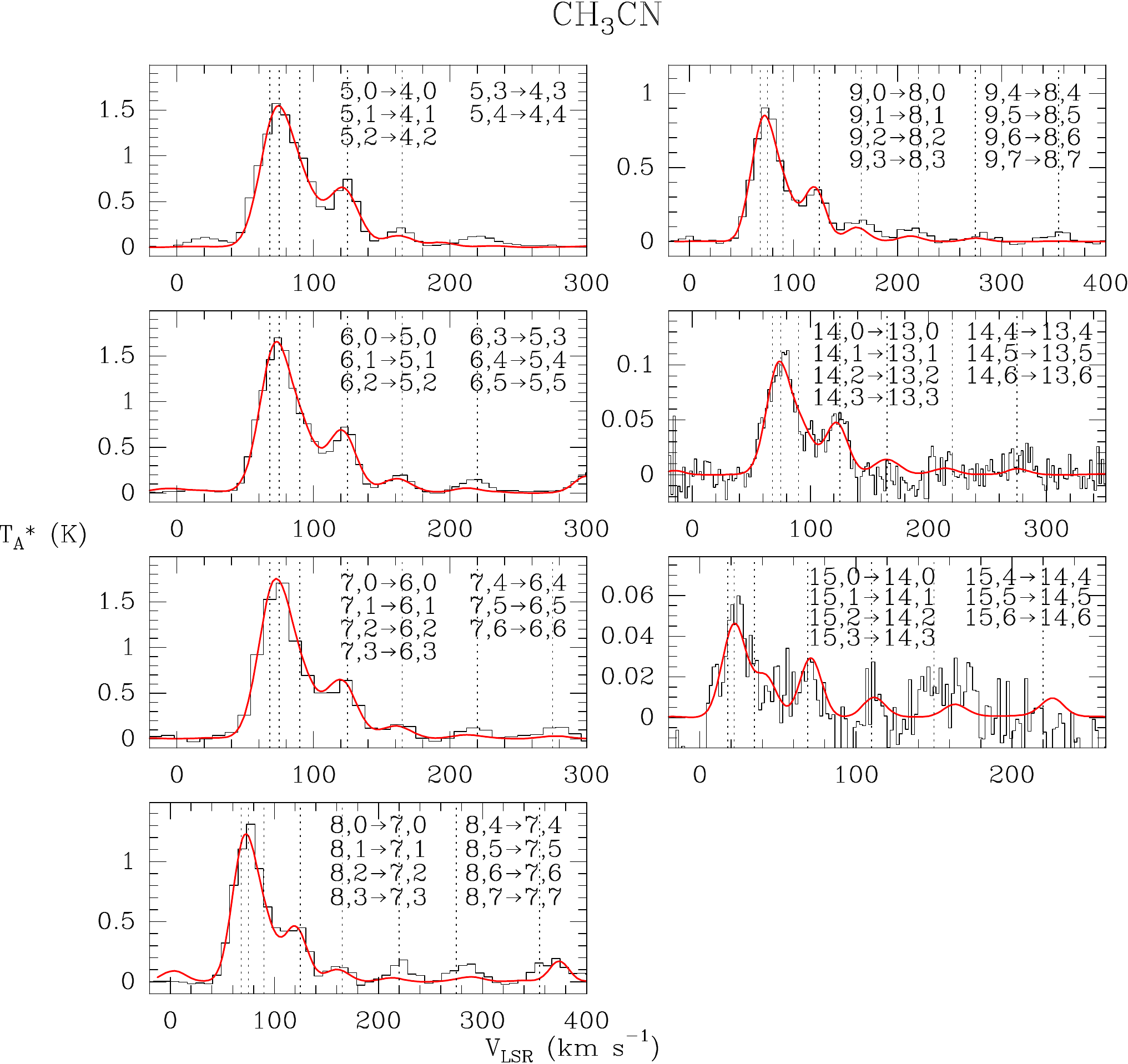}
 \caption{CH$_3$CN detected lines (black histogram spectra). The LTE best fits from MADCUBA are also shown overlaid in red. Each panel from top to bottom, from left to right presents the detected K-ladder for J=5$\rightarrow$4, J=6$\rightarrow$5, J=7$\rightarrow$6, J=8$\rightarrow$7, J=9$\rightarrow$8, J=14$\rightarrow$13, and J=15$\rightarrow$14 transitions respectively. The vertical dotted lines indicate detected each K-transition within the same J-transition. Note that other fitted lines that are not aligned with the dotted lines are arose from other species.}
    \label{fig-CH3CN}
\end{figure*}
%%%%%%%%%%%%%%%%%%%%%%%%%%%%%%%%%%%%%%%%%%
\begin{figure*}
\includegraphics[width = \textwidth]{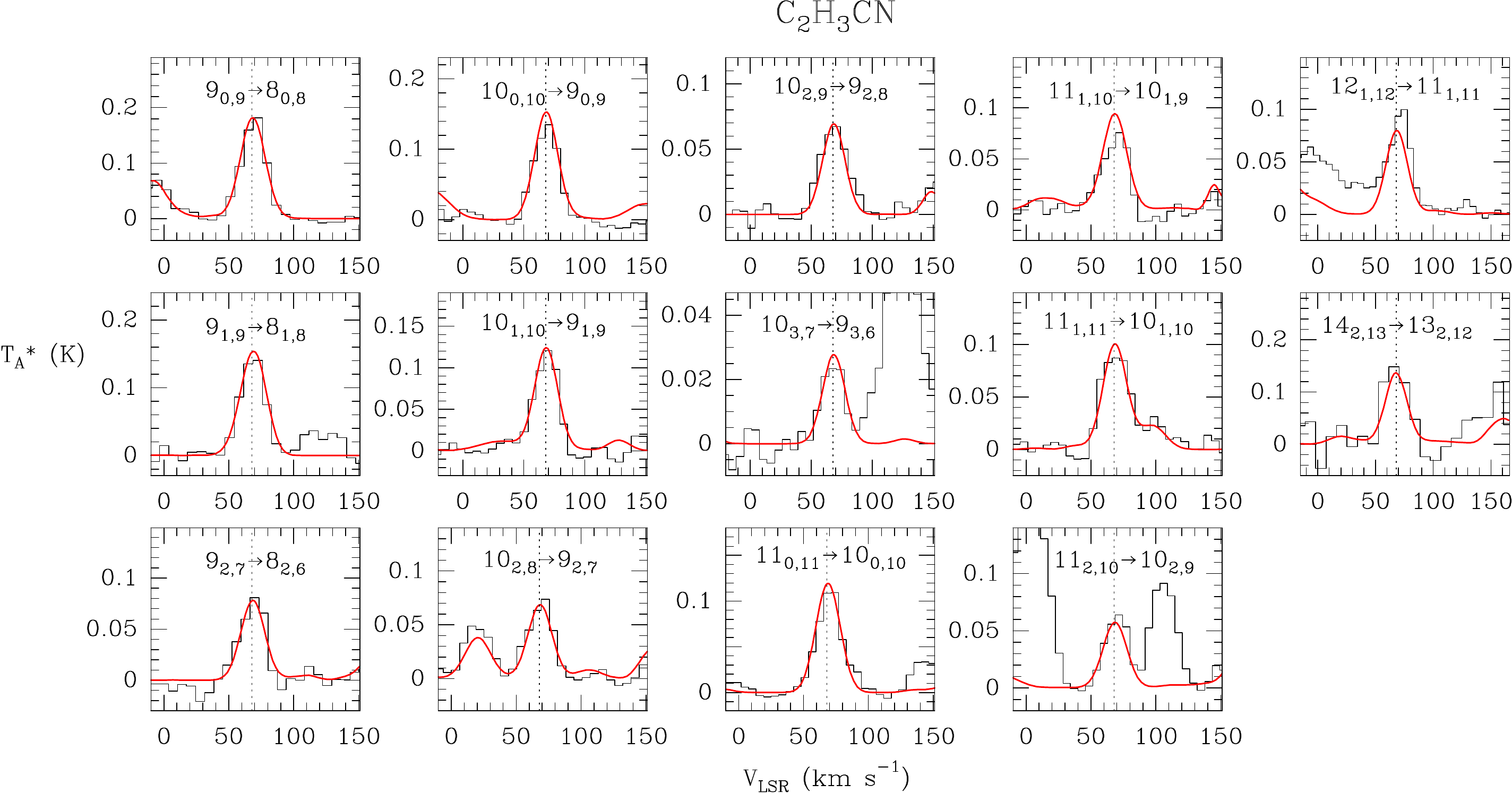}
 \caption{C$_2$H$_3$CN detected lines (black histogram spectra). The LTE best fits from MADCUBA are also shown overlaid in red. The vertical dotted lines correspond to the central velocity of the G+0.693 cloud. Note that other fitted lines that are not aligned with the dotted lines are arose from other species.}
    \label{fig-C2H3CN}
\end{figure*}
%%%%%%%%%%%%%%%%%%%%%%%%%%%%%%%%%%%%%%%%%%
\begin{figure*}
\includegraphics[width = 0.85\textwidth]{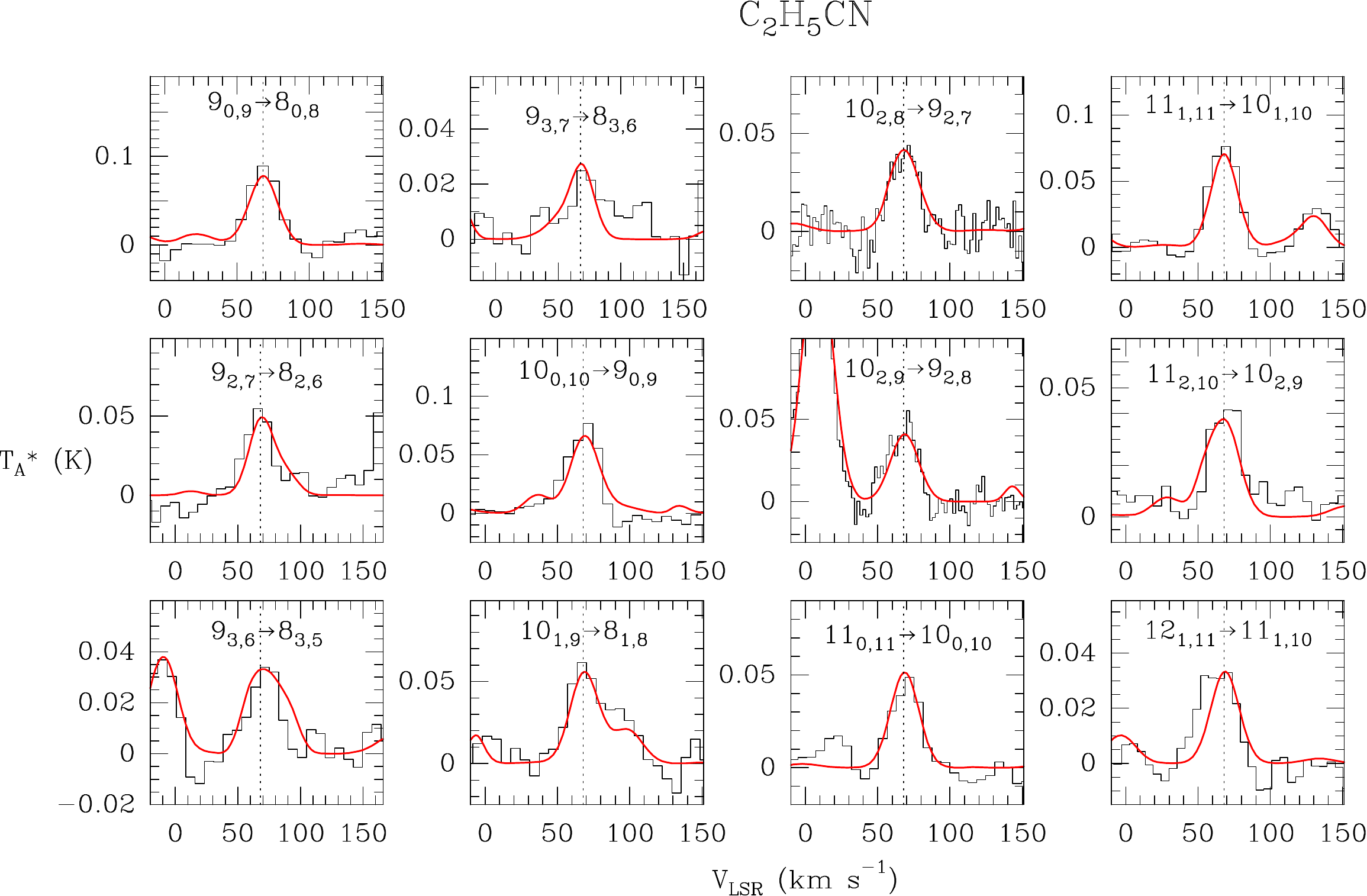}
 \caption{C$_2$H$_5$CN detected lines (black histogram spectra). The LTE best fits from MADCUBA are also shown overlaid in red. The vertical dotted lines correspond to the central velocity of the G+0.693 cloud. Note that other fitted lines that are not aligned with the dotted lines are arose from other species.}
    \label{fig-C2H5CN}
\end{figure*}
%%%%%%%%%%%%%%%%%%%%%%%%%%%%%%%%%%%%%%%%%%
\begin{figure*}
\includegraphics[width = 0.6\textwidth]{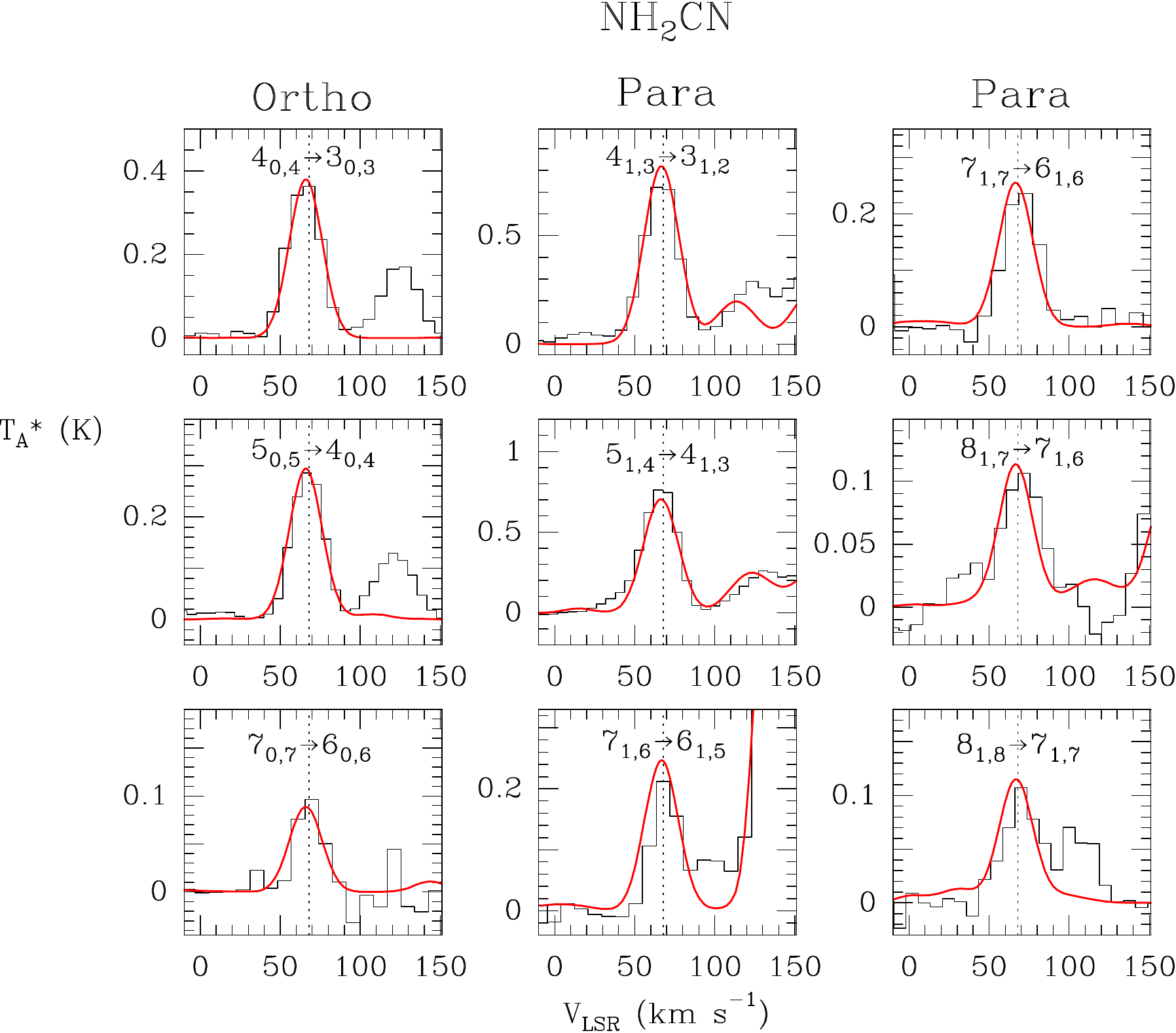}
 \caption{NH$_2$CN detected lines (black histogram spectra).The LTE best fits from MADCUBA are also shown overlaid in red. The vertical dotted lines correspond to the central velocity of the G+0.693 cloud. Note that other fitted lines that are not aligned with the dotted lines are arose from other species.}
    \label{fig-NH2CN}
\end{figure*}
%%%%%%%%%%%%%%%%%%%%%%%%%%%%%%%%%%%%%%%%%%
\begin{figure*}
\includegraphics[width = \textwidth]{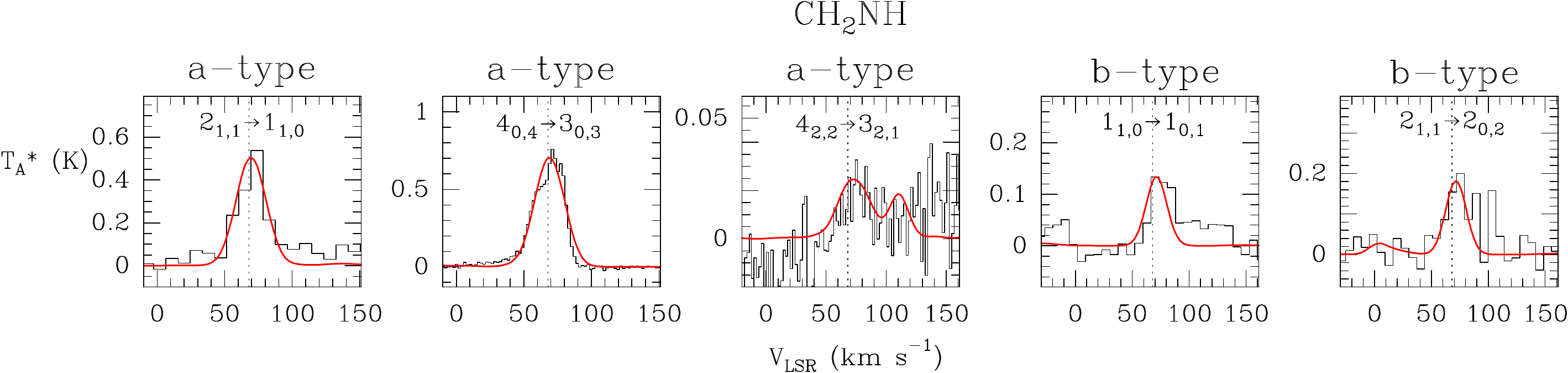}
 \caption{CH$_2$NH detected lines (black histogram spectra). The LTE best fits from MADCUBA are also shown overlaid in red. The vertical dotted lines correspond to the central velocity of the G+0.693 cloud. Note that other fitted lines that are not aligned with the dotted lines are arose from other species.}
    \label{fig-CH2NH}
\end{figure*}
%%%%%%%%%%%%%%%%%%%%%%%%%%%%%%%%%%%%%%%%%%
\begin{figure*}
\includegraphics[width = 0.8\textwidth]{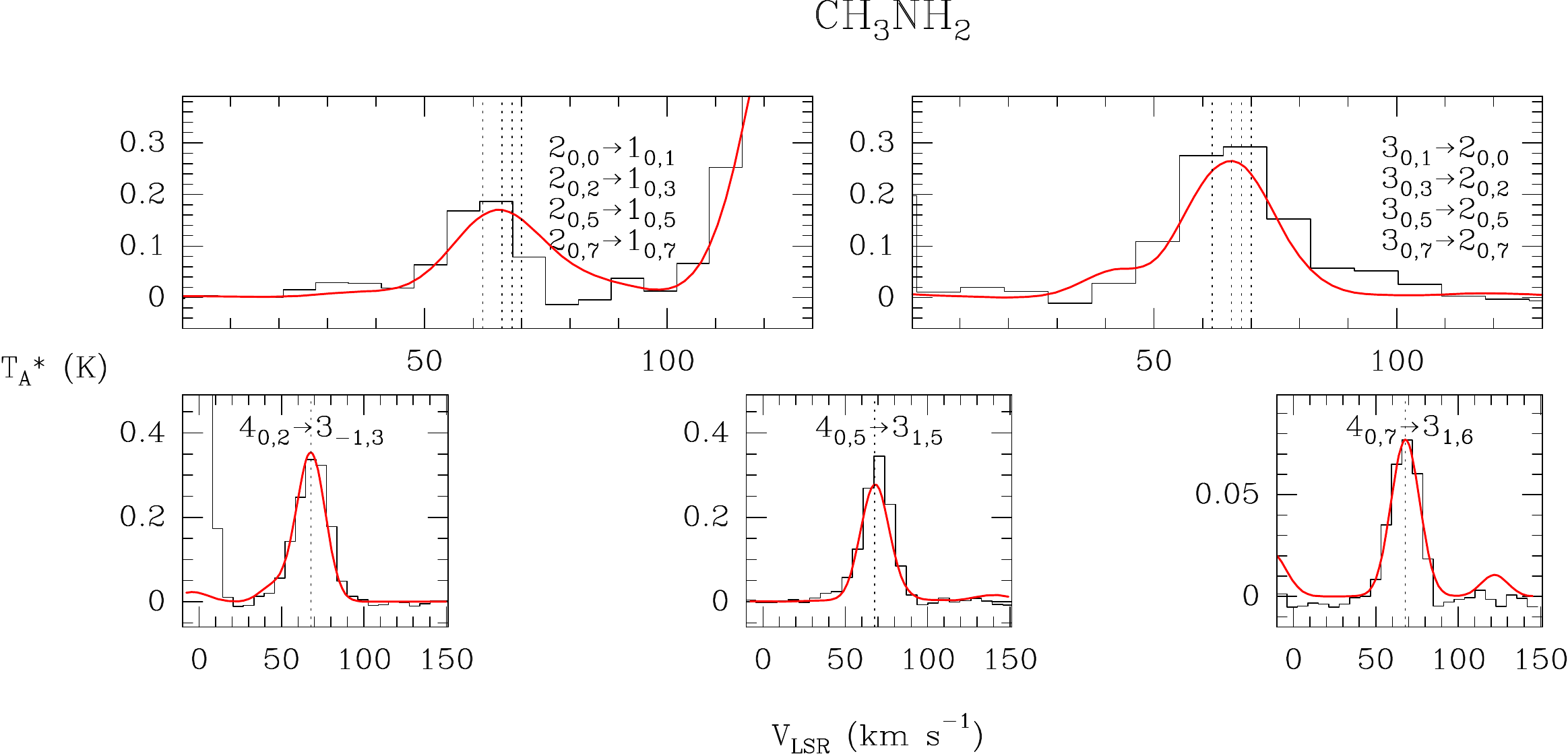}
 \caption{CH$_3$NH$_2$ detected lines (black histogram spectra).The LTE best fits from MADCUBA are also shown overlaid in red. Top panel present the blended hyperfine splittings of J=2$\rightarrow$1 and J=3$\rightarrow$2 transition. The bottom panel presents resolved hyperfine components of J=4$\rightarrow$3 transition.}
    \label{fig-CH3NH2}
\end{figure*}
%%%%%%%%%%%%%%%%%%%%%%%%%%%%%%%%%%%%%%%%%%
\begin{figure*}
\includegraphics[width = \textwidth]{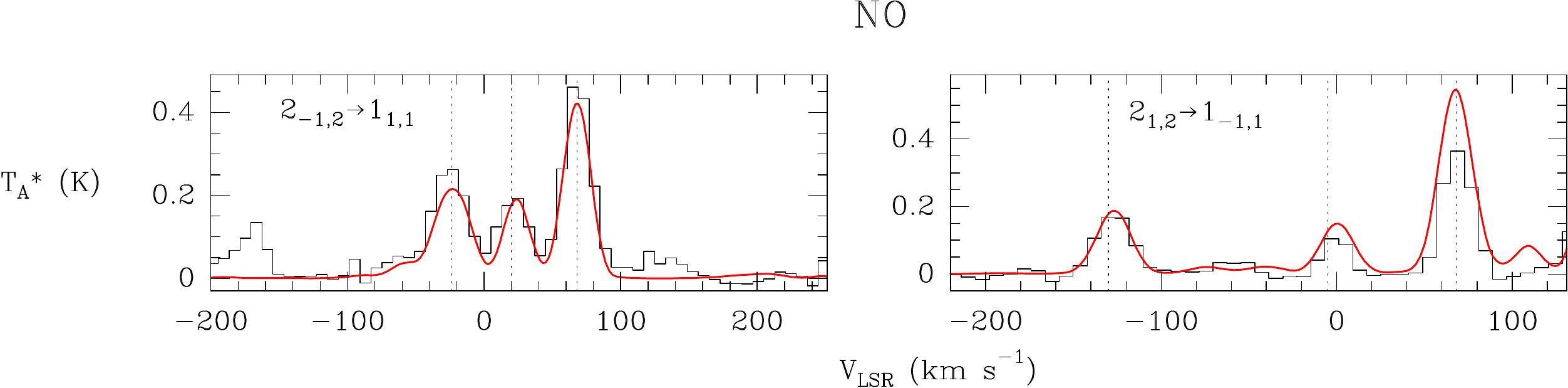}
 \caption{NO detected lines (black histogram spectra). The LTE best fits from MADCUBA are also shown overlaid in red. Dotted lines indicate the detected hyperfine components of each transition.}
    \label{fig-NO}
\end{figure*}
%%%%%%%%%%%%%%%%%%%%%%%%%%%%%%%%%%%%%%%%%%
\begin{figure*}
\includegraphics[width = 0.85\textwidth]{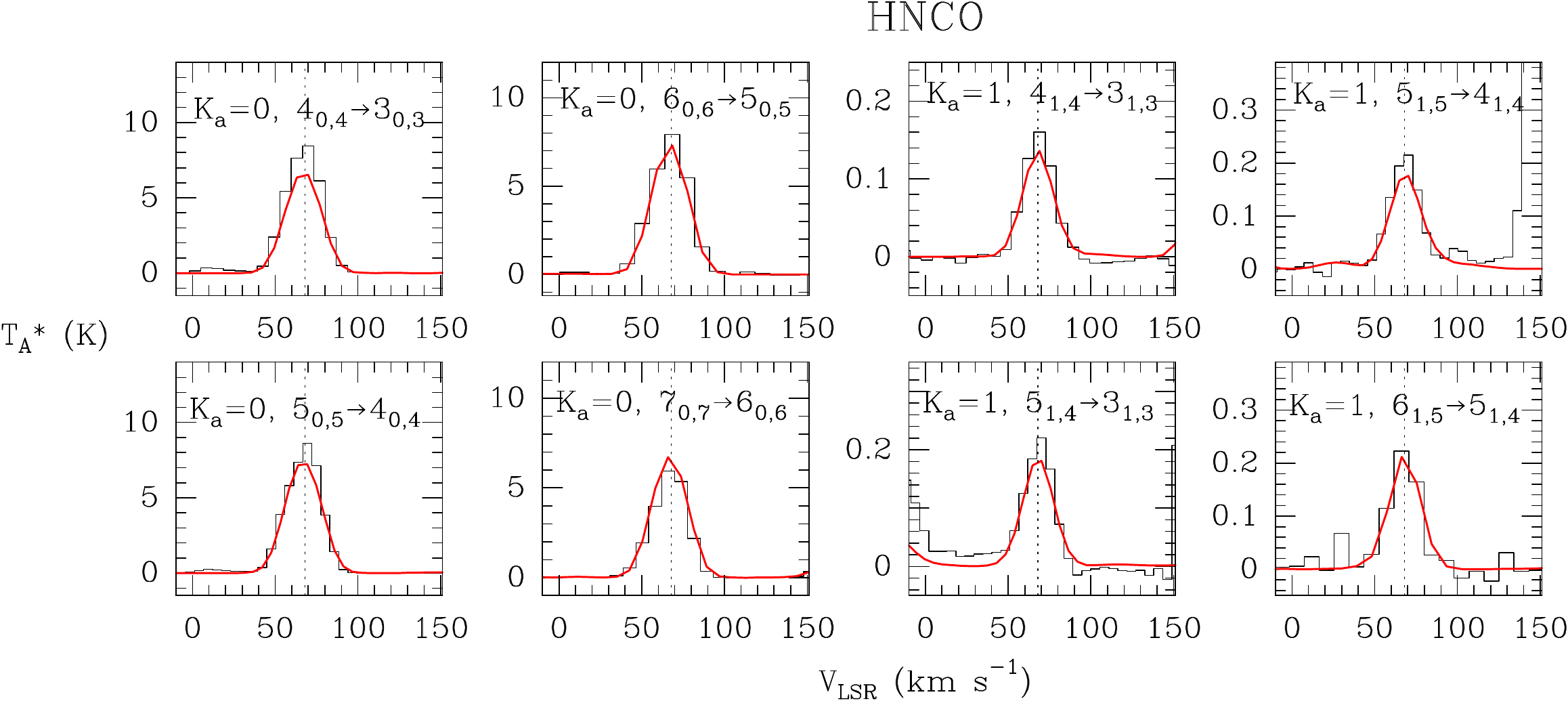}
 \caption{HNCO detected lines (black histogram spectra). The LTE best fits from MADCUBA are also shown overlaid in red. The vertical dotted lines correspond to the central velocity of the G+0.693 cloud.}
    \label{fig-HNCO}
\end{figure*}
%%%%%%%%%%%%%%%%%%%%%%%%%%%%%%%%%%%%%%%%%%
\begin{figure*}
\includegraphics[width = 0.85\textwidth]{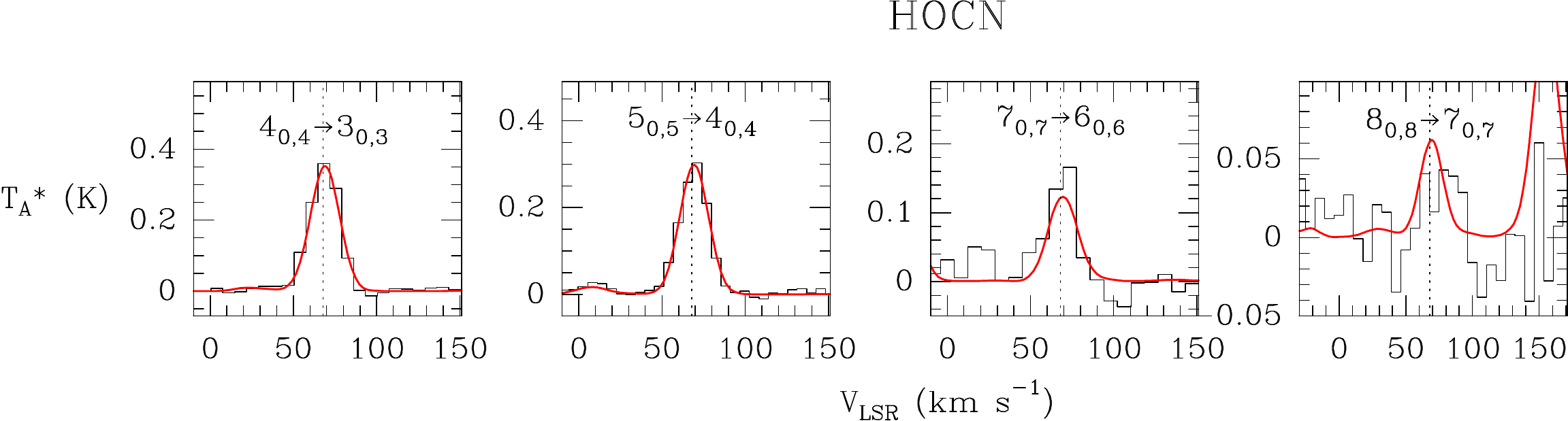}
 \caption{HOCN detected lines (black histogram spectra). The LTE best fits from MADCUBA are also shown overlaid in red. The vertical dotted lines correspond to the central velocity of the G+0.693 cloud. Note that other fitted lines that are not aligned with the dotted lines are arose from other species.}
    \label{fig-HOCN}
\end{figure*}
%%%%%%%%%%%%%%%%%%%%%%%%%%%%%%%%%%%%%%%%%%
\begin{figure*}
\includegraphics[width = 0.85\textwidth]{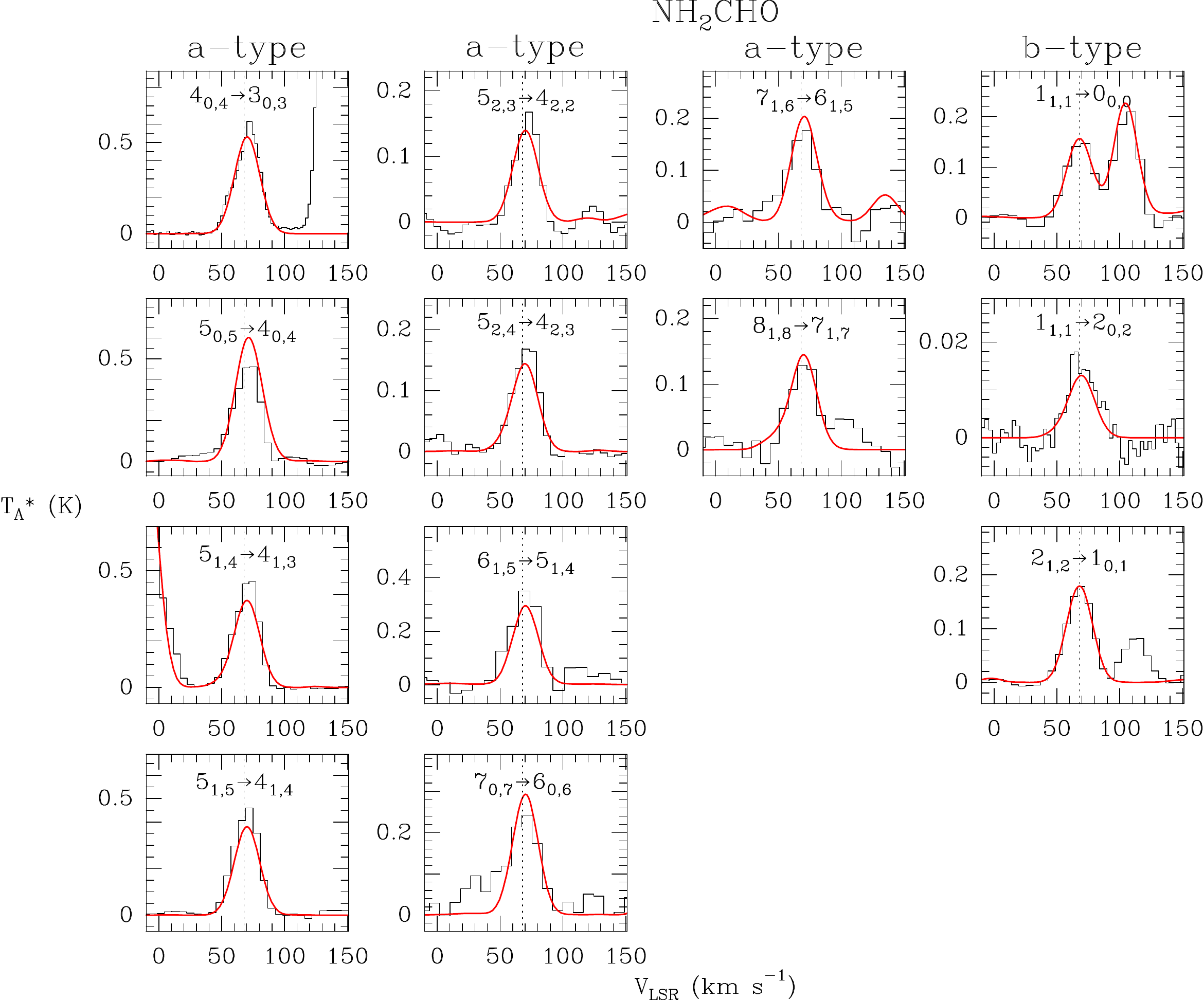}
 \caption{NH$_2$CHO detected lines (black histogram spectra). The LTE best fits from MADCUBA are also shown overlaid in red. The vertical dotted lines correspond to the central velocity of the G+0.693 cloud. Note that other fitted lines that are not aligned with the dotted lines are arose from other species.}
    \label{fig-NH2CHO}
\end{figure*}
%%%%%%%%%%%%%%%%%%%%%%%%%%%%%%%%%%%%%%%%%%
\begin{figure*}
\includegraphics[width = 0.7\textwidth]{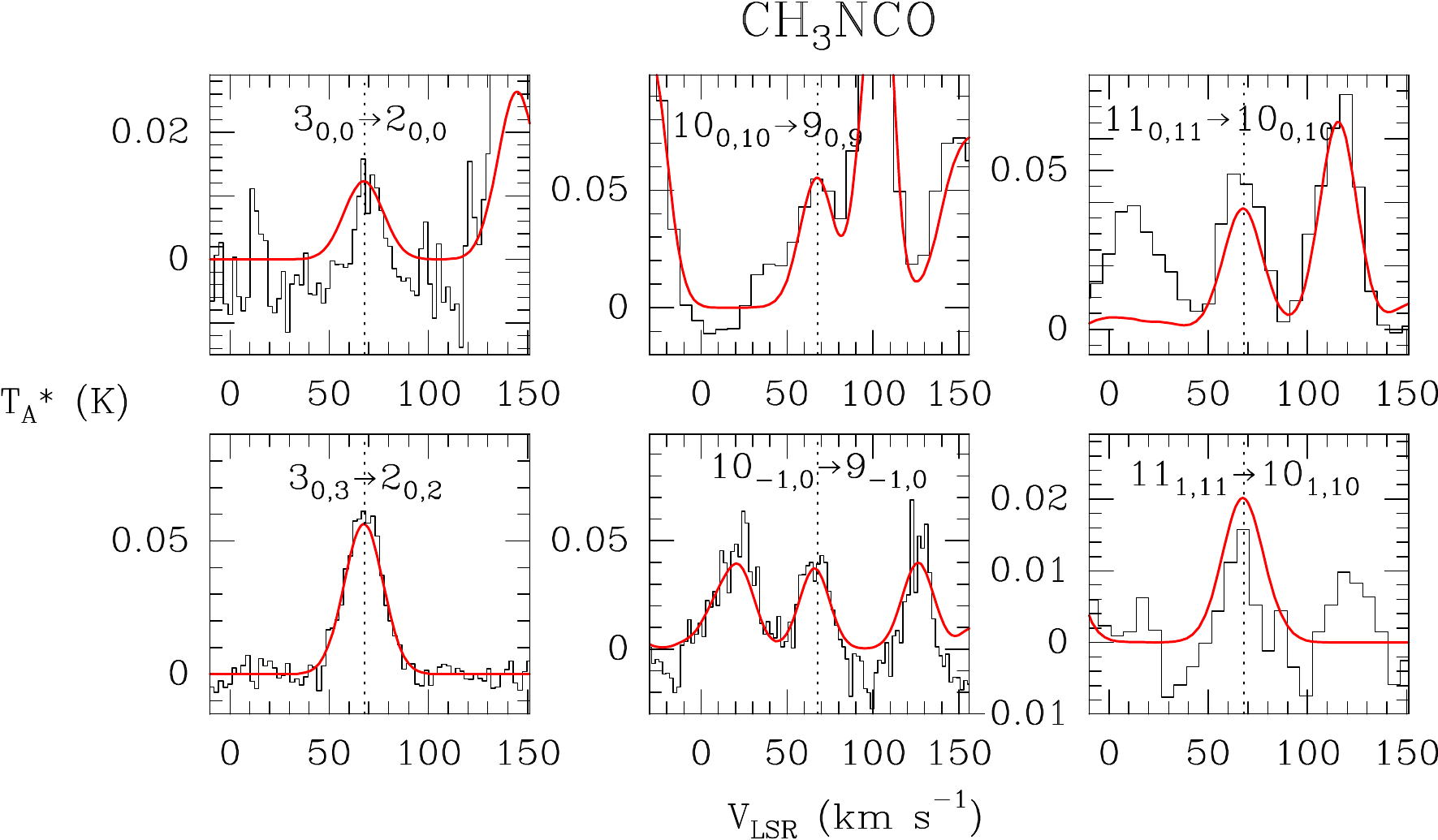}
 \caption{CH$_3$NCO detected lines (black histogram spectra). The LTE best fits from MADCUBA are also shown overlaid in red. The vertical dotted lines correspond to the central velocity of the G+0.693 cloud. Note that other fitted lines that are not aligned with the dotted lines are arose from other species.}
    \label{fig-CH3NCO}
\end{figure*}

%%%%%%%%%%%%%%%%%%%%%%%%%%%%%%%%%%%%%%%%%%
%%%%%%%% FIGURE-FITTING %%%%%%%%%%%%%%%%
%%%%%%%%%%%%%%%%%%%%%%%%%%%%%%%%%%%%%%%%%%

% Don't change these lines
\bsp	% typesetting comment
\label{lastpage}
\end{document}